\documentclass[%
superscriptaddress,
notitlepage,
showpacs,
amsmath,amssymb,
aps,
longbibliography,
]{revtex4-2}

\usepackage{psfrag,graphicx,epsfig,color}
\usepackage{dcolumn}
\usepackage{bm}
\usepackage{natbib}
\usepackage{float}
\usepackage[usenames,dvipsnames,svgnames,table]{xcolor}
\usepackage{subfigure}
\usepackage{nicefrac}

\def\re    {{R_\lambda}}
\def\uu {{\mathbf{u}}}

\def\ww {{\boldsymbol{\omega}}}

\def\SS {{\mathbf{S}}}

\def\HH {{\mathbf{H}}}

\def\AA {{\mathbf{A}}}

\definecolor{mygreen}{rgb}{0,0.7,0.}

\begin{document}

\title{
Role of pressure in generation of intense
velocity gradients in turbulent flows
}

\author{Dhawal Buaria}
\email[]{dhawal.buaria@nyu.edu}
\affiliation{Tandon School of Engineering, New York University, New York, NY 11201, USA}
\affiliation{Max Planck Institute for Dynamics and Self-Organization, 37077 G\"ottingen, Germany}
\author{Alain Pumir}
\affiliation{Ecole Normale Superieure, Universite de Lyon 1 and CNRS, 69007 Lyon, France}
\affiliation{Max Planck Institute for Dynamics and Self-Organization, 37077 G\"ottingen, Germany}

\date{\today}

\begin{abstract}

We investigate the role of pressure, via its Hessian tensor $\HH$, on amplification of vorticity and strain-rate and contrast it with other inviscid nonlinear mechanisms. Results are obtained from direct numerical simulations of isotropic turbulence with Taylor-scale Reynolds number in the range $140-1300$. Decomposing $\HH$ into local isotropic ($\HH^{\rm I}$) and nonlocal deviatoric  ($\HH^{\rm D}$) components reveals that $\HH^{\rm I}$ depletes vortex stretching (VS), whereas $\HH^{\rm D}$ enables it, with the former slightly stronger. The resulting inhibition is significantly weaker than the nonlinear mechanism which always enables VS. However, in regions of intense vorticity, identified using conditional statistics, contribution from $\HH$ dominates over nonlinearity, leading to overall depletion of VS. We also observe near-perfect alignment between vorticity and the eigenvector of $\HH$ corresponding to the smallest eigenvalue, which conforms with well-known vortex-tubes. We discuss the connection between this depletion, essentially due to (local) $\HH^{\rm I}$, and recently identified self-attenuation mechanism [Buaria et al. {\em Nat. Commun.} 11:5852 (2020)], whereby intense vorticity is locally attenuated through inviscid effects. In contrast, the influence of $\HH$ on strain-amplification is weak. It opposes strain self-amplification, together with VS, but its effect is much weaker than VS. Correspondingly, the eigenvectors of strain and $\HH$ do not exhibit any strong alignments. For all results, the dependence on Reynolds number is very weak. In addition to the fundamental insights, our work provides useful data and validation benchmarks for future modeling endeavors, for instance in Lagrangian modeling of velocity gradient dynamics, where conditional $\HH$ is explicitly modeled.

\end{abstract}

\maketitle


\section{Introduction}

A defining feature of turbulent flows
is the generation of small scale structures, 
leading to dramatic enhancement of 
transport rates of mass, momentum and 
energy. In this regard, the importance
of statistical properties of the velocity gradient tensor
$\AA = \nabla \uu$, where $\uu$ is the velocity 
field, is well recognized
\cite{Frisch95, Sreeni97, falkovich01, Wallace09, Tsi2009, Meneveau11}. 
By taking the derivative of the incompressible
Navier-Stokes equations, the dynamics of $\AA$ are given 
by the following transport equation 
\begin{align}
\frac{D A_{ij}}{Dt} = - A_{ik} A_{kj} - H_{ij} + \nu \nabla^2 A_{ij}
\label{eq:dadt}
\end{align}
where $D/Dt$ is the material derivative, $\nu$ is the kinematic
viscosity,  
$H_{ij} = \partial^2 P/\partial x_i \partial x_j$ is the pressure Hessian tensor;
and the incompressibility condition imposes $A_{ii} = 0$. 
The quadratic non-linearity in Eq.~\eqref{eq:dadt} 
captures the self-amplification of velocity gradients,
which leads to intermittent generation of extreme events
and small scale structures in the flow.
Owing to their practical significance in 
various physical processes
\cite{falkovich02, hamlington12, BSY.2015, voth2017, BCSY2021a}, 
their postulated universality
\cite{K41a, Frisch95, Sreeni97},  as well as their
connection to potential singularities of Euler and 
Navier-Stokes equations \cite{gibbon:2008, doering2009, Fefferman}, 
the study of small scales and velocity gradients
is of obvious importance
in turbulence theory and modeling.


As implied by Eq.~\eqref{eq:dadt}, the dynamics 
of velocity gradients are influenced by the
pressure Hessian tensor. This leads to a nonlocal coupling of the 
entire gradient field, since pressure satisfies
the Poisson equation: 
$\nabla^2 P = - A_{ij} A_{ji}$, as obtained
by taking the trace of Eq.~\eqref{eq:dadt}. 
The mathematical difficulties 
posed by this nonlocality 
makes it very hard to decipher the precise role of pressure Hessian 
on gradient amplification 
and the formation of extreme events. 
In general, it has been observed that 
pressure Hessian acts to counteract
the nonlinear amplification 
\cite{nomura:1998, Tsi2009, kalelkar2006, Lawson2015, Carbone:20b, BPB2022}.
Since the pressure Hessian is a symmetric tensor, 
its influence on the amplification of strain, the
symmetric part of the $\AA$, is more explicit
\cite{nomura:1998, BPB2020}. 
In contrast, its influence on vorticity, 
the skew-symmetric part of $\AA$, 
is indirectly felt through strain,
and much more difficult to understand.
The interaction between vorticity and strain itself
is an indispensable ingredient of turbulence.
For instance, it is well established that
vorticity preferentially aligns with the eigenvector corresponding 
to the intermediate eigenvalue of the strain tensor, 
which in turn is positive on average \cite{Ashurst87, Tsi2009}.
Additionally, this alignment is considerably stronger
in regions of intense vorticity 
and strain \cite{BBP2020, BPB2022}. 
In contrast, the role of pressure Hessian, 
especially in regions of intense vorticity or strain 
has received little or no attention.

Prior studies predominantly 
focused on unconditional statistics,
which do not distinguish 
quiescent regions from regions where extreme events reside. 
Additionally, 
they have also been restricted to low Reynolds numbers
\cite{nomura:1998, tsi99, Lawson2015}. 
It is well known that extreme vorticity and strain
events in turbulence have a pronounced structure,
where the statistical properties can be very different
than the mean field
\cite{Jimenez93, moisy:2004, Tsi2009, BBP2020, BPB2022}.
Thus, analyzing statistics conditioned on the magnitude
of vorticity or strain can be particularly useful
to understand the underlying amplification mechanism
\cite{Tsi2009, BBP2020, BPB2022}.
In addition to providing fundamental insights,
conditional statistics of pressure Hessian 
also play a central role in
turbulence modeling, particularly
Lagrangian modeling of velocity
gradient dynamics
\cite{girimaji:1990, Meneveau11, Lawson2015, JM.arfm, tian21, bs_pnas_23}.

In this work, our objective is to
systematically analyze the effect of pressure
Hessian on amplification of  
vorticity and strain. 
We identify and analyze various 
correlations between the pressure
Hessian and vorticity and strain fields.
We consider unconditional statistics
and also statistics conditioned on
magnitude of vorticity and strain, 
to focus on extreme events.
It is worth noting that strain itself can be nonlocally
related to vorticity, via the Biot-Savart integral,
providing an alternative way to study
the nonlocality of gradient amplification---without
invoking pressure---by filtering strain into scale-wise contributions
\cite{ham_pof08, BPB2020, BP2021}.
Alternatively, the pressure Hessian tensor itself can also be 
filtered into scale-wise 
contributions \cite{Vlaykov2019Small}.
Complementary to these approaches, our focus here is to directly 
analyze the pressure Hessian term to 
directly investigate its role on gradient amplification.

To that end, the necessary statistics are extracted from
state-of-the-art direct numerical
simulations of isotropic turbulence 
in periodic domains, which is the most
efficient numerical tool to study the small-scale properties 
of turbulence. One important purpose of the current study is also 
to understand the effect of increasing Reynolds number. 
To this end, we utilize a massive DNS database with Taylor-scale 
Reynolds number $\re$ ranging from $140$
to $1300$, on up to grid sizes of $12288^3$; 
particular attention is given on having good
small-scale resolution to accurately resolve the extreme events
\cite{BPBY2019, BBP2020, BPB2022}, for which conditional statistics
are analyzed.


The manuscript is organized as follows. 
The necessary background for our analysis is briefly reviewed in \S~2. 
The numerical approach and DNS database
is presented in \S~3.  In \S~4, the role
of pressure Hessian is analyzed in the context
of vorticity amplification, whereas
in \S~5 the analysis is
in the context of strain amplification.
Finally, we summarize our results in
\S~5.

\section{Background}

The vorticity vector $\ww$ and the strain tensor $\SS$,
defined as: 
$\omega_i=\varepsilon_{ijk} A_{jk}$
($\varepsilon_{ijk}$ being the Levi-Civita symbol)
and $S_{ij} = (A_{ij} + A_{ji})/2$, respectively
represent the skew-symmetric and symmetric components
of the velocity gradient
tensor, and characterize the local rotational and
stretching motions.
Their evolution equations can be readily obtained
from Eq.~\eqref{eq:dadt} and are given as
\begin{align}
\label{eq:ws_vrt} 
\frac{D \omega_i}{Dt}  &=  \omega_j S_{ij} + \nu \nabla^2 \omega_i \ ,  \\
\frac{D S_{ij}}{Dt} &= -S_{ik} S_{kj} - \frac{1}{4} (\omega_i \omega_j 
- \omega^2 \delta_{ij}) 
- H_{ij} + \nu \nabla^2 S_{ij}  \ .
\label{eq:ws_str}
\end{align}
The non-linear amplification of vorticity is captured by
the vortex stretching vector $W_i = \omega_j S_{ij}$,
whereas amplification of strain is controlled by the 
self-amplification term and additionally via feedback of vorticity. 
Although the pressure Hessian, which is
a symmetric tensor, 
only contributes to evolution of strain,
it still indirectly affects vorticity,
since the pressure Poisson couples both 
vorticty and strain. Indeed, taking
the trace of Eq.~\eqref{eq:ws_str}, gives
$\nabla^2 P = (\omega_i \omega_i - 2 S_{ij} S_{ij})/2$.
The influence of pressure Hessian on vorticity 
becomes apparent when considering
the  evolution equation for the vortex
stretching vector
\begin{align}
\frac{D W_i}{Dt} = 
- \omega_j H_{ij} + \text{viscous terms} \ , 
\label{eq:w1}
\end{align}
Note that $D W_i/Dt = D^2 \omega_i /Dt^2$ in the inviscid
limit, which directly relates the pressure Hessian 
to the second derivative of vorticity.


To quantify the intensity of gradients,
we consider the magnitudes
of vorticity and strain \cite{BPBY2019, BP2022}
\begin{align}
    \Omega = \omega_i \omega_i \ , \ \ \  \Sigma = 2 S_{ij} S_{ij} \ ,
\end{align}
where the former is the enstrophy, and the latter is dissipation rate $\epsilon$
divided by viscosity, i.e, $\Sigma = \epsilon / \nu$. 
In homogeneous turbulence,
$\langle \Omega \rangle = \langle \Sigma \rangle = 1/\tau_K^2 $, where
$\tau_K$ is the Kolmogorov time scale. Likewise, it is useful
to consider transport equations for these quantities
\begin{align}
\label{eq:OS_vrt}
\frac{1}{2} \frac{D \Omega }{Dt}  &=  \omega_i \omega_j S_{ij} + \nu \omega_i \nabla^2 \omega_i \ ,  \\
\frac{1}{4} \frac{D \Sigma}{Dt} &= -S_{ij} S_{jk} S_{ki} 
- \frac{1}{4} \omega_i \omega_j S_{ij}
- S_{ij} H_{ij} + \nu S_{ij} \nabla^2 S_{ij} 
\label{eq:OS_str}
\end{align}
The amplification of enstrophy is engendered by the term 
$\omega_i \omega_j S_{ij} = \omega_i W_i$, which in turn evolves
according to:
\begin{align}
\frac{D \omega_i W_i }{Dt}  =
W_i W_i -  \omega_i \omega_j H_{ij} + \text{viscous terms} \  . 
\label{eq:w2}
\end{align}
In the inviscid limit, $D (\omega_i W_i)/Dt = D^2 \Omega /Dt^2$, thus
Eq.~\eqref{eq:w2} is complementary to 
Eq.~\eqref{eq:w1} for $W_i$. 
The above equations identify the correlations
responsible for generation of intense velocity gradients,
which we will analyze, both unconditionally 
and conditioned on magnitudes of $\Omega$ and $\Sigma$.

For our analysis, it is also useful to consider 
the eigenframes of the strain and pressure Hessian tensors. 
For the strain tensor, it is defined by the eigenvalues
$\lambda_i$ (for $i=1,2,3$), such that $\lambda_1 \ge \lambda_2 \ge \lambda_3$
and the corresponding eigenvectors $\mathbf{e}_i$. 
Incompressibility gives
$\lambda_1 + \lambda_2 + \lambda_3=0$, implying 
that $\lambda_1$ is always positive (stretching) and $\lambda_3$ is
always negative (compressive).  
Similarly, the eigenframe of pressure Hessian
is given by the eigenvalues $\lambda_i^{\rm P}$ (in descending order), 
and eigenvectors $\mathbf{e}_i^{\rm P}$. 
In this case, incompressibility gives 
$H_{ii} = \nabla^2 P = \lambda_1^{\rm P} + \lambda_2^{\rm P} + \lambda_3^{\rm P}$,
which is in general non-zero. Thus, 
it is convenient to decompose the pressure Hessian 
into isotropic and deviatoric components: 
\begin{align}
    \label{eq:HI}
H_{ij}^{\rm I}  &=  \frac{H_{kk}}{3} \delta_{ij} \ ,  
\ \ \ \text{with} \ \  H_{kk} = \nabla^2 P = (\Omega - \Sigma)/2  \ , \\ 
H_{ij}^{\rm D} &= H_{ij} - H_{ij}^{\rm I}  \  .
    \label{eq:HD}
\end{align}
Note, since $\HH^{\rm I}$ can be explicitly expressed in terms
of $\AA$, it can be considered local, whereas
$\HH^{\rm D}$ captures the nonlocality of the pressure field \cite{Ohkitani:95}.
The eigenvalues $\lambda_i^{\rm D}$ of $H_{ij}^{\rm D}$ satisfy 
$\lambda_i^{\rm D} = \lambda_i^{\rm P} - H_{kk}/3$ and hence,
$\lambda_1^{\rm D} + \lambda_2^{\rm D} + \lambda_3^{\rm D} = 0$,
implying $\lambda_1^{\rm D} > $ and $\lambda_3^{\rm D} <0$; 
whereas the eigenvectors are unaffected, i.e.,
$\mathbf{e}_i^{\rm D} = \mathbf{e}_i^{\rm P}$.
Using this framework, we obtain
\begin{align}
    \omega_i \omega_j H_{ij} = \lambda_i^{\rm P} (\mathbf{e}_i^{\rm P} \cdot \ww)^2 \ , 
    \ \
    \omega_i \omega_j H_{ij}^{\rm D} = \lambda_i^{\rm D} (\mathbf{e}_i^{\rm P} \cdot \ww)^2 \ ,
    \ \ 
    \omega_i \omega_j H_{ij}^{\rm I} = \Omega (\Omega - \Sigma)/6  \ , 
\label{eq:decomp}
\end{align}
which decomposes the correlation between 
vorticity and pressure Hessian into individual contribution from each eigendirection.
Similarly, the terms $\omega_i W_i$ and $W_i W_i$ can be
decomposed in the eigenframe of the strain tensor, illustrating
the importance of 
the alignments between vorticity vector and the eigenvectors of strain.
We refer to our recent works \cite{BBP2020, BPB2022}
for a discussion of these properties, 
On the other hand, the contribution
$S_{ij} H_{ij}$ in Eq.~\eqref{eq:OS_str}
can be written as 
\begin{align}
 S_{ij} H_{ij} =  \lambda_i^{\rm P} \lambda_j 
 (\mathbf{e}_i^{\rm P} \cdot \mathbf{e}_j )^2  \ . 
\label{eq:sh}
\end{align}
Since, $ S_{ij} H_{ij}^{\rm I} = 0$ 
from incompressibility, it also follows that
\begin{align}
 S_{ij} H_{ij} =   S_{ij} H_{ij}^{\rm D} = 
 \lambda_i^{\rm D} \lambda_j 
 (\mathbf{e}_i^{\rm P} \cdot \mathbf{e}_j )^2  \ . 
\label{eq:sh_inc}
\end{align}

\section{Numerical approach and database}

\begin{table}
\begin{center}
\begin{tabular}{lccccc}
    $\re$   & $N^3$    & $k_{max}\eta$ & $T_E/\tau_K$ & $T_{\rm sim}$ & $N_s$  \\
\hline
    140 & $1024^3$ & 5.82 & 16.0 & 6.5$T_E$ &  24 \\
    240 & $2048^3$ & 5.70 & 30.3 & 6.0$T_E$ &  24 \\
    390 & $4096^3$ & 5.81 & 48.4 & 2.8$T_E$ &  35 \\
    650 & $8192^3$ & 5.65 & 74.4 & 2.0$T_E$ &  40 \\
   1300 & $12288^3$ & 2.95 & 147.4 & 20$\tau_K$ &  18 \\
\hline
\end{tabular}
\caption{Simulation parameters for the DNS runs
used in the current work: 
the Taylor-scale Reynolds number ($\re$),
the number of grid points ($N^3$),
spatial resolution ($k_{max}\eta$), 
ratio of large-eddy turnover time ($T_E$)
to Kolmogorov time scale ($\tau_K$),
length of simulation ($T_{\rm sim}$) in statistically stationary state
and the number of instantaneous snapshots ($N_s$) 
used for each run to obtain the statistics.
}
\label{tab:param}
\end{center}
\end{table}

The data utilized here are the same
as in several recent works 
\cite{BS2020, BBP2020, BPB2020, BP2021, BS2022, BS2023}
and are generated using
direct numerical simulations (DNS) of incompressible
Navier-Stokes equations, for the canonical setup
of isotropic turbulence in a periodic
domain. The simulations are carried out
using highly accurate Fourier pseudo-spectral methods
with second-order Runge-Kutta integration in time,
and the large scales are forced numerically
to achieve statistical stationarity \cite{Rogallo}.
A key characteristic  of our data is that we have achieved
a wide range of Taylor-scale Reynolds number $\re$,
going from $140-1300$, while maintaining excellent small-scale
resolution, which is as high as
$k_{\rm max} \eta \approx 6$, where
$k_{\rm max} = \sqrt{2}N/3$, is the maximum
resolved wavenumber on a $N^3$ grid, and $\eta$
is the Kolmogorov length scale.
Convergence with respect to resolution and statistical sampling has been
adequately established in previous works.
For convenience, we summarize the DNS database and the simulation
parameters in Table~\ref{tab:param}; additional
details can be obtained in our prior works cited earlier.

\section{Role of pressure Hessian on vorticity amplification}
\label{sec:vort_hess}

\subsection{Unconditional statistics}
\label{subsec:uncond}

Table~\ref{tab:un} lists various unconditional statistics
characterizing the role of pressure Hessian on vortex stretching, 
based on Eq.~\eqref{eq:w2}, at different $\re$. 
All quantities are appropriately non-dimensionalized by
the Kolmogorov time scale $\tau_K$, and henceforth, should be 
interpreted as such (unless otherwise mentioned). 
We first consider the eigenvalues of  pressure Hessian.
Owing to homogeneity $\langle H_{ii} \rangle= 0$;
thus, $\sum_{i=1}^{3} \langle \lambda_i^{\rm P} \rangle = 0$ and
$\langle \lambda_i^{\rm P} \rangle = \langle \lambda_i^{\rm D} \rangle$. 
Table~\ref{tab:un} reveals that the individual averages of 
the eigenvalues are approximately equal to $0.3: 0.03: -0.33$, 
without any appreciable dependence on $\re$.
The intermediate eigenvalue is overall positive on average but its magnitude
is substantially smaller than the other two, and essentially
close to zero.

\begin{table}
\begin{tabular}{l|c|c|ccc|c}
    $\re$ & $\langle \lambda_i^{\rm P} \rangle$ & $\langle (\mathbf{e}_i^{\rm P} \cdot \hat{\ww})^2\rangle$ &  $\langle \omega_i \omega_j H_{ij}^{\rm D} \rangle$  & $\langle \omega_i \omega_j H_{ij}^{\rm I} \rangle$ & $\langle \omega_i \omega_j H_{ij} \rangle$ & $\langle W_i W_i \rangle$ \\
\hline    
    140 & 0.295 : 0.024 : -0.319 & 0.290 : 0.416 : 0.294 & -0.363 & 0.410 & 0.047 & 0.146 \\
    240 & 0.301 : 0.025 : -0.326 & 0.292 : 0.418 : 0.290 & -0.483 & 0.549 & 0.066 & 0.180 \\
    390 & 0.302 : 0.025 : -0.327 & 0.292 : 0.418 : 0.290 & -0.587 & 0.670 & 0.083 & 0.216 \\
    650 & 0.304 : 0.026 : -0.330 & 0.293 : 0.418 : 0.289 & -0.746 & 0.853 & 0.107 & 0.269 \\
   1300 & 0.305 : 0.026 : -0.331 & 0.293 : 0.418 : 0.289 & -1.010 & 1.153 & 0.146 & 0.360 \\
\end{tabular}
\caption{
Unconditional averages of various quantities associated
with correlation of vorticity and pressure Hessian,
based on Eq.~\eqref{eq:w2}.
$\lambda_i^{\rm P}$, for $i=1,2,3$, are the eigenvalues of 
pressure Hessian, with corresponding eigenvectors
$\mathbf{e}_i^{\rm P}$. All quantities are appropriately normalized 
by the Kolmogorov time scale $\tau_K$. 
}
\label{tab:un}
\end{table}

Similarly, 
the mean square of alignment cosines between vorticity 
and eigenvectors of pressure Hessian, 
$ \langle (\mathbf{e}_i^{\rm P} \cdot \hat{\ww})^2 \rangle$, 
where $\hat{\ww}$ denotes the  unit vector parallel to $\ww$,  
are also essentially independent of $\re$.  
Note that the square of cosines sum up to unity
for all three directions. Additionally, they are 
bounded between 0 and 1 
for each individual direction, respectively for the
case of perfect orthogonal and parallel alignment;
whereas for a uniform distribution of the alignment cosine,
the mean square average is $1/3$. 
From Table~\ref{tab:un}, we observe that
the measured alignments do not deviate significantly 
from $1/3$, 
with a weak preferential alignment of vorticity
with the intermediate eigenvector
of the pressure Hessian.

\begin{figure}
\centering
\includegraphics[width=0.55\linewidth]{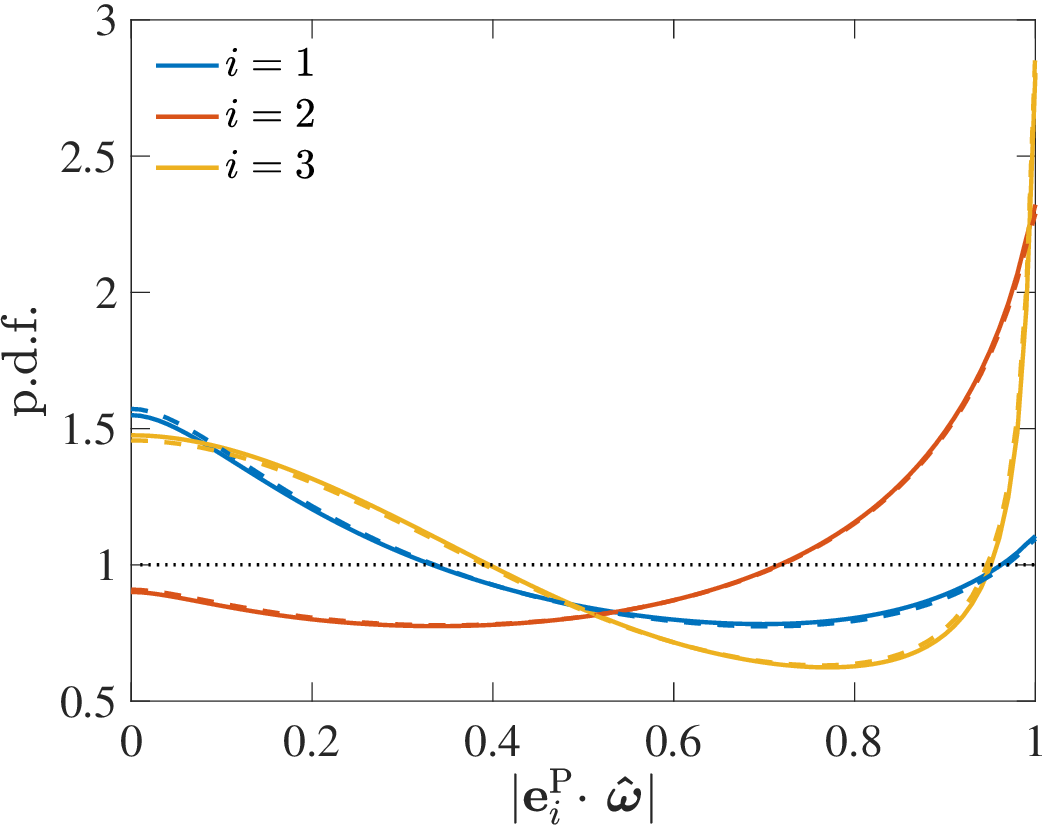}
\caption{Probability density function (p.d.f.) of
alignment cosines between vorticity 
and eigenvectors of pressure Hessian,
at $\re=1300$ (solid lines) and $\re=140$ (dashed lines). 
}
\label{fig:align_wp}
\end{figure}

It is worth noting that while a uniform distribution of the alignment
cosine implies the second moment is $1/3$, the reverse is not
necessarily true.  Thus, it is useful to also inspect the
probability density function (p.d.f.) of the alignment cosines,
which are shown in Fig.~\ref{fig:align}. The solid and dashed curves
respectively correspond to $\re=1300$ and $140$, demonstrating
that the alignments are independent of $\re$. 
The distributions for 
$ |\mathbf{e}_1^{\rm P} \cdot \hat{\ww}|$
and $ |\mathbf{e}_1^{\rm P} \cdot \hat{\ww}|$
conform with expectation from their second moments in
Table~\ref{tab:un}, respectively indicating weak
preferential orthogonal and parallel alignment. 
However, for $ |\mathbf{e}_3^{\rm P} \cdot \hat{\ww}|$, we
observe an anomalous behavior, showing
simultaneous preferential orthogonal and parallel alignments,
which cancel each other out when evaluating the second moment. 
Note that the limiting cases of 0 or 1 for the second
moment of alignment cosines do not present such an anomaly
for the PDFs. 
In fact, we will see later that when 
considering conditional statistics,
the alignments 
are substantially enhanced for extreme vorticity events
(such that using the second moment only does not lead to any ambiguity).

Finally, we consider the net contributions to the budget
of vortex stretching (as per Eq.~\eqref{eq:w2}). 
Table~\ref{tab:un} shows separately the mean contributions 
from the deviatoric and isotropic components of the pressure Hessian
and their sum $\langle \omega_i \omega_j H_{ij} \rangle $, 
contrasted with the nonlinear term $ \langle W_i W_i \rangle $.
Remarkably, the  deviatoric and isotropic contributions,
are comparable in magnitude but opposite in sign.
Taking into account the negative sign before the pressure Hessian term in Eq.\eqref{eq:w2}, 
it follows that $\HH^{\rm D}$ favors vortex stretching, 
whereas $\HH^{\rm I}$ inhibits it.
This essentially establishes 
that nonlocal effects of the pressure
field enable vortex stretching \cite{Ohkitani:95},   
conforming with earlier
studies that vortex stretching is predominantly nonlocal
\cite{ham_pof08, BPB2020, BP2021}.
We will further 
elaborate on this effect later. 


The strong cancellation between the deviatoric
and isotropic contributions results in a weakly
positive value of
$\langle \omega_i \omega_j H_{ij} \rangle $,
implying that pressure Hessian overall weakly opposes 
vortex stretching -- which is primarily a local effect
stemming from the dominant isotropic contribution of
pressure Hessian. 
Both the deviatoric and isotropic contributions become
stronger with $\re$, but the latter is always slightly 
dominant.
In contrast, the term $W_i W_i$, listed last in Table~\ref{tab:un}, 
is positive by definition, and thus always enables vortex stretching. 
This term increases
with $\re$ and is  noticeably  
larger than $\langle \omega_i \omega_j H_{ij} \rangle$
(at all $\re$). This leads to
the (anticipated) result
that the nonlinear
effects in turbulence predominantly enable vortex stretching,
with a net positive energy cascade from large to small scales
\cite{batchelor53, Betchov56, kerr85}. 
Although not shown, 
the $\re$ dependence of these quantities 
conforms with an approximate power law 
of $R_\lambda^{0.39}$ in nominal
agreement with fourth order moment
of velocity derivatives \cite{gylfason:2004, BS_PRL_2022}.

\subsection{Conditional statistics}  

In the previous subsection, we considered
unconditional statistics, which provide an overall
perspective of the flow. 
To specifically characterize the extreme events 
or regions of intense vorticity, we condition
the statistics on the magnitude of vorticity;
specifically, we will use $\Omega\tau_K^2$,
(or equivalently, $\Omega/\langle \Omega \rangle $), 
to quantify the extremeness of an event with respect to the
mean field.

\begin{figure}
\centering
\includegraphics[width=0.55\linewidth]{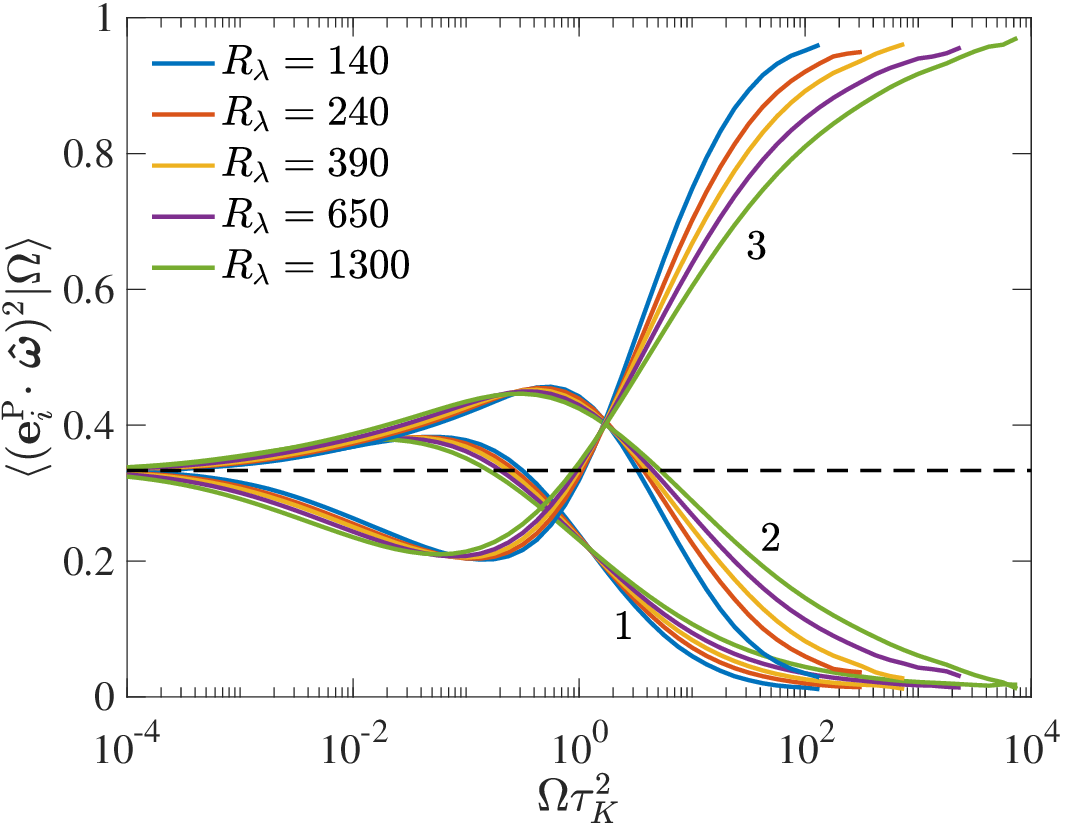}
\caption{
Conditional expectation (given enstrophy $\Omega$) 
of second moment of alignment cosines between vorticity
unit vector ($\hat{\ww}$)
and eigenvectors of pressure Hessian ($\mathbf{e}_i^{\rm P}$),
at various $\re$. 
}
\label{fig:align}
\end{figure}

Figure~\ref{fig:align} shows the conditional mean square of 
alignment cosines between vorticity and the eigenvectors of pressure 
Hessian, at various $\re$.
For weak enstrophy, 
all the curves are at $1/3$,
consistent with a uniform distribution of cosines. 
For $\Omega \tau_K^2 \approx 1$, 
we notice that vorticity has a weak preferential alignment
with $\mathbf{e}^{\rm P}_2$, in agreement with 
the unconditional results in Table~\ref{tab:un}. 
However, for extreme events, a  different picture emerges
and vorticity almost perfectly aligns with $\mathbf{e}^{\rm P}_3$
(becoming orthogonal to both first and second eigenvectors).
This alignment can be explained by considering
the familiar picture of intense vorticity being 
arranged in tube-like structures
\cite{Jimenez93, Ishihara07, BPBY2019}, as 
discussed at the end of the present subsection.

\begin{figure}
\centering
\includegraphics[width=0.485\linewidth]{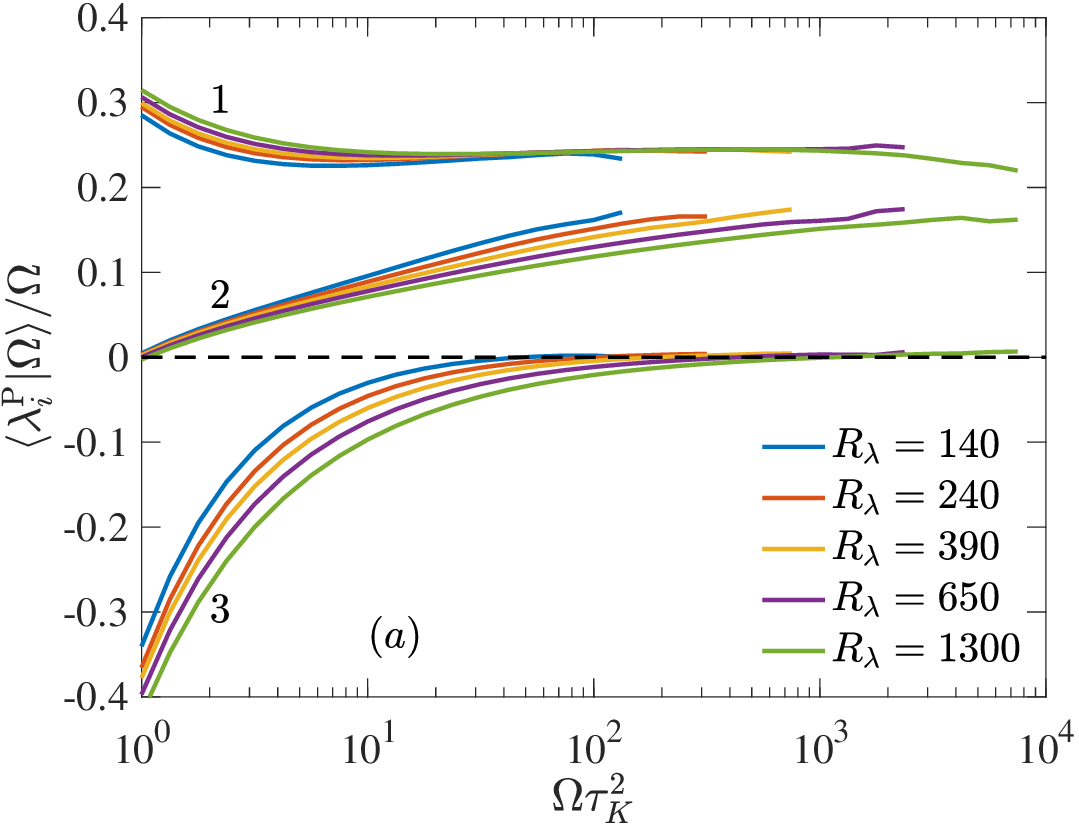} \ \ \
\includegraphics[width=0.485\linewidth]{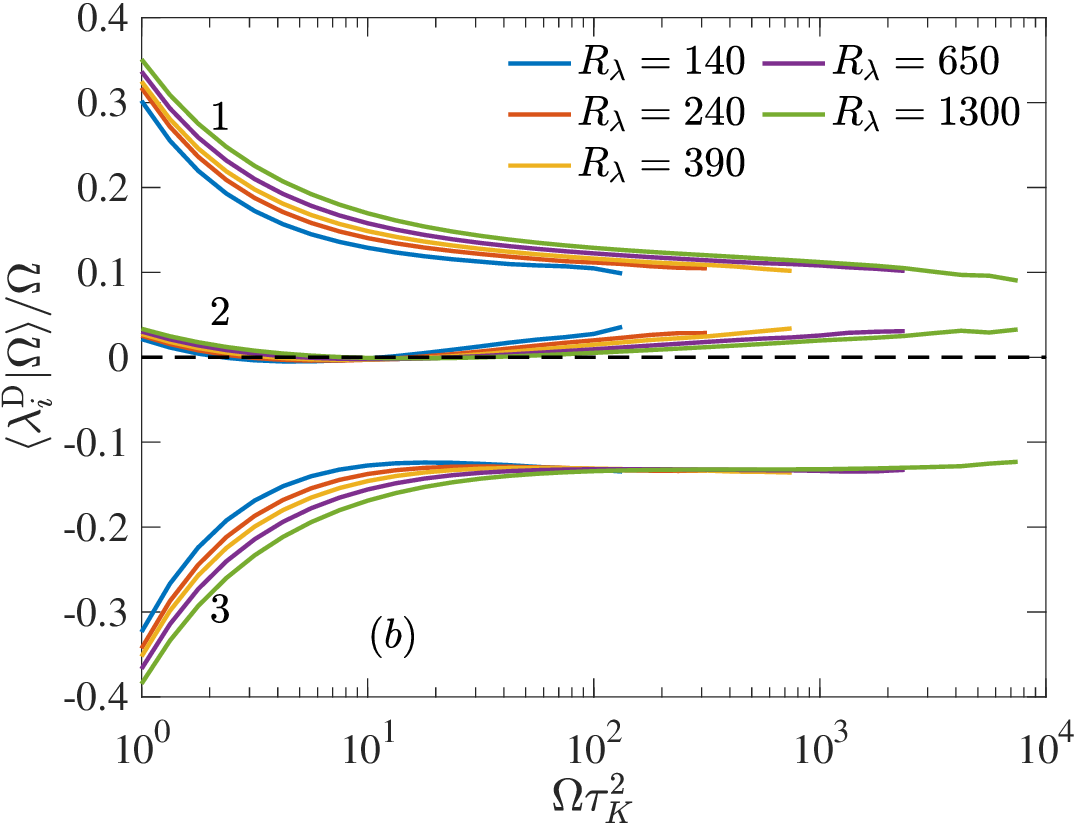} 
\caption{
(a) Conditional expectations (given enstrophy $\Omega$) of the 
eigenvalues of pressure Hessian, for various $\re$.
(b) Conditional expectations of the eigenvalues of
the deviatoric part. 
The legend in panel (b) also applies to (a). 
}
\label{fig:lam1}
\end{figure}

Figure~\ref{fig:lam1} shows the conditional expectation 
of the eigenvalues of pressure Hessian. 
To focus on extreme events, the abscissa 
is set to $\Omega \tau_K^2 \ge 1$.
The identity $\sum_i \lambda_i^{\rm P} = \nabla^2 P = (\Omega - \Sigma)/2$ 
implies that $\sum_i \langle \lambda_i^{\rm P} | \Omega \rangle  =  
(\Omega - \langle \Sigma|\Omega \rangle)/2$.
Recent numerical works \cite{BPBY2019, BBP2020, BP2022} 
have shown that for large enstrophy,
$\langle \Sigma |\Omega  \rangle \sim \Omega^\gamma$,
where the exponent $\gamma < 1$, but weakly
increases with $\re$
(with $\gamma \to 1$ being the asymptotic limit). 
Thus, while the average sum of eigenvalues is zero for the mean field, 
it is strongly positive in regions of intense enstrophy. 
In fact, to the leading order, one can anticipate
that $\langle \lambda_i^{\rm P} | \Omega \rangle \sim \Omega$ when 
$\Omega \tau_K^2 \gg 1$.
Figure~\ref{fig:lam1}a shows the quantity 
$\langle \lambda_i^{\rm P} | \Omega \rangle / \Omega$.
The behavior of the averaged 
eigenvalues at 
$\Omega \tau_K^2 =1$ is
consistent with the results in Table~\ref{tab:un}. 
However, in regions of intense enstrophy, the first two eigenvalues 
are strongly
positive, whereas the third eigenvalue is essentially zero. 
Once again, no appreciable dependence on $\re$ is seen for this case.


\begin{figure}
\centering
\includegraphics[width=0.485\linewidth]{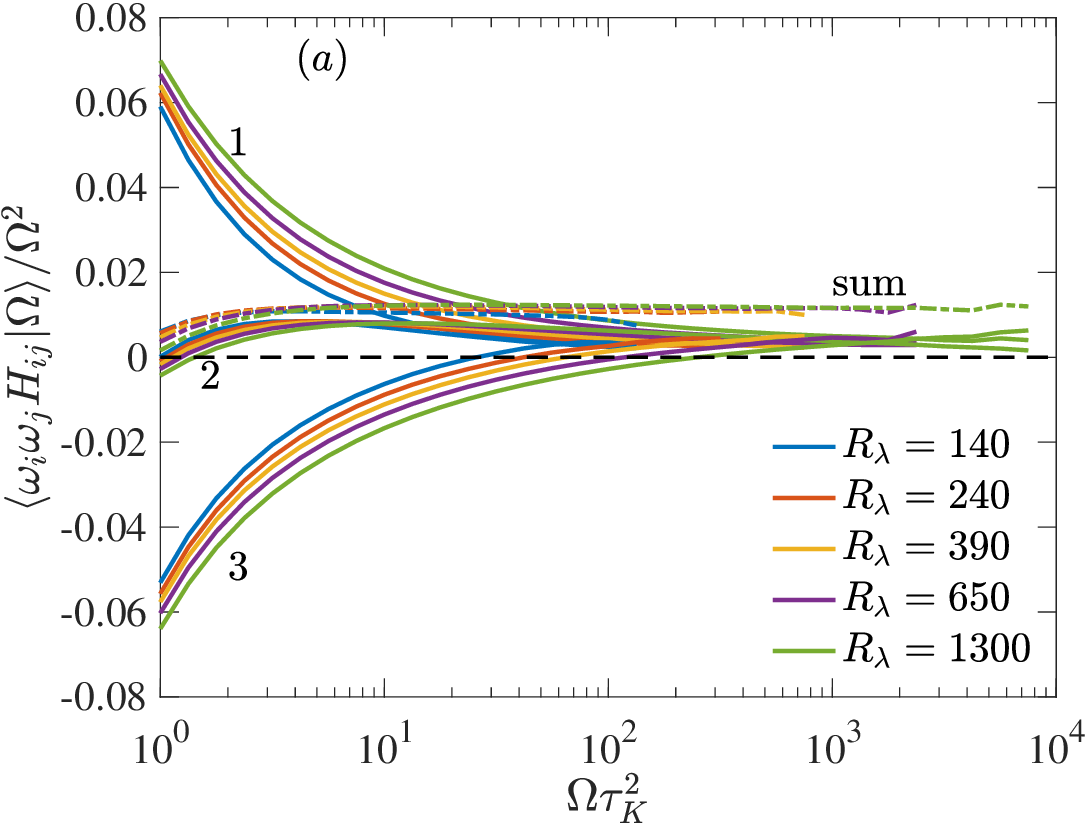} \\
\vspace{0.5cm}
\includegraphics[width=0.485\linewidth]{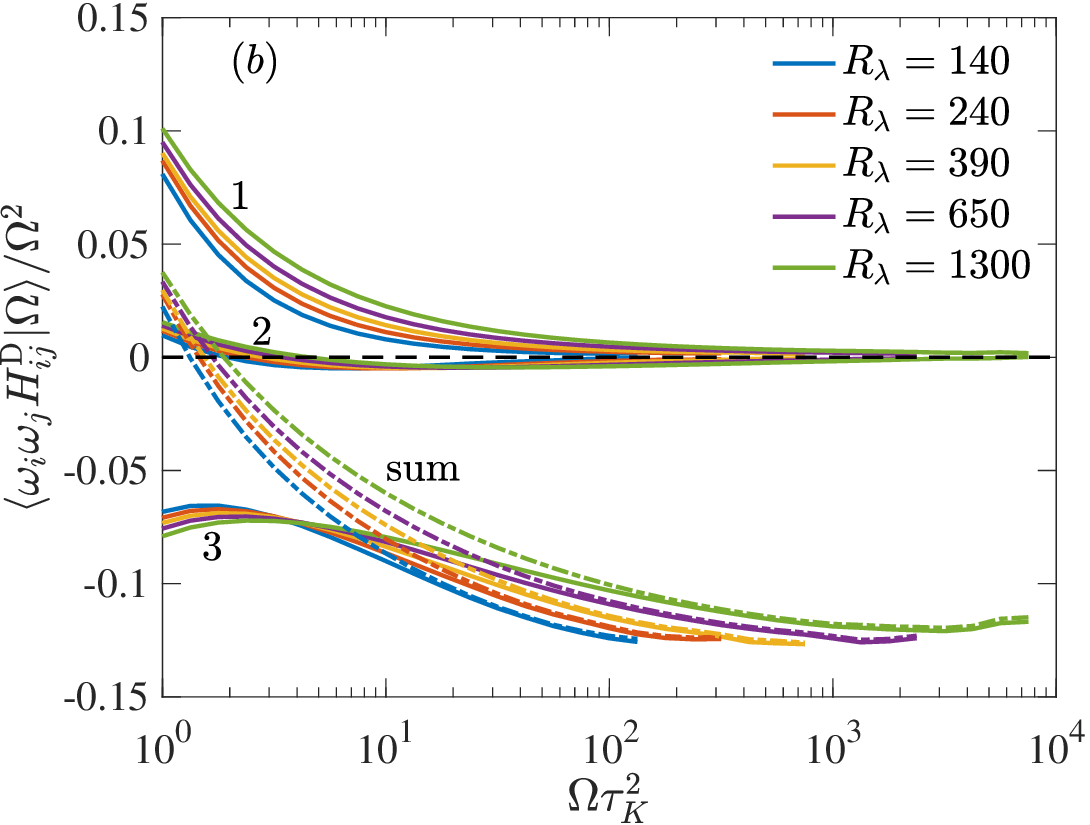} \ \ \ 
\includegraphics[width=0.485\linewidth]{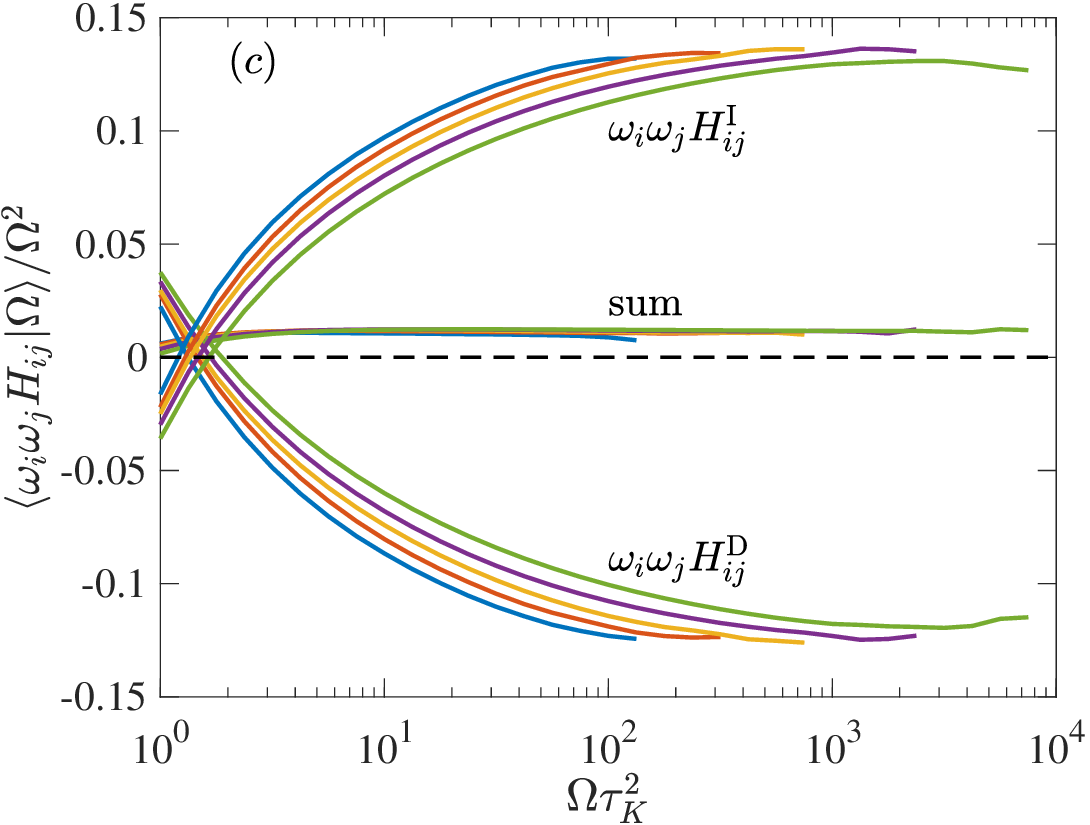} 
\caption{
(a) Conditional expectation (given enstrophy $\Omega$) 
of the term $\omega_i \omega_j H_{ij}$ (marked as sum)
and individual contributions from each eigendirection 
of pressure Hessian (in solid lines). 
The quantities are normalized by $\Omega^2$ to reveal 
a plateau like behavior for $\Omega \tau_K^2 \gg 1 $. 
(b) Same as a, but for $\omega_i \omega_j H_{ij}^{\rm D}$,
(the contribution from the deviatoric part of pressure Hessian). 
(c) The net contributions from a and b are further contrasted
with $\omega_i \omega_j H_{ij}^{\rm I}$ (the contribution
from isotropic part of pressure Hessian).  
}
\label{fig:lam2}
\end{figure}

Given that vorticity is almost perfectly aligned with third
eigenvector of pressure Hessian
and the corresponding
eigenvalue is close to zero,  
the behavior of 
the term $\omega_i \omega_j H_{ij}$ cannot be easily inferred.
Figure~\ref{fig:lam2}a shows the
conditional expectation  
$\langle \omega_i \omega_j H_{ij} |\Omega \rangle/\Omega^2$, along 
with the individual contributions
in each eigendirection. The normalization 
by $\Omega^2$ comes from simple dimensional 
grounds. 
For $\Omega\tau_K^2 \simeq 1$,
the contributions from first and third eigendirections
are expectedly positive and negative, respectively,
largely canceling each other to give a
weakly positive net contribution 
(whereas the contribution from second direction 
is essentially negligible).
However, for $\Omega\tau_K^2 \gg 1$,
the contributions from all directions
become positive and comparable.
Thus, in regions of intense vorticity,
the role of pressure (Hessian)
is to oppose vortex stretching.

To better understand this result, 
we consider next the contributions from the deviatoric and 
the isotropic components of the pressure Hessian.
The sum of eigenvalues 
of the deviatoric part $\HH^{\rm D}$
is always constrained to 
be zero, i.e., $\sum_i \lambda_i^D = 0$, 
and thus,   
$\sum_i \langle \lambda_i^{\rm D} | \Omega \rangle = 0$.
Figure~\ref{fig:lam1}b shows the conditional expectation
$\langle \lambda_i^{\rm D} | \Omega \rangle / \Omega$.
While the results shown in Fig.~\ref{fig:lam1}b for $\Omega \tau_K^2 \simeq 1$
are essentially identical to 
$\langle \lambda_i^P | \Omega \rangle$,
the behavior for $\Omega \tau_K \gg 1$ is
different, with the first
and second eigenvalues being always positive
(with the second being noticeably weaker),
whereas third eigenvalue is always negative
(perfectly canceling the contribution from other
two).  Thus, from the results 
in Figs.\ref{fig:align} and \ref{fig:lam1}b, 
it can be  anticipated
that $\langle \omega_i \omega_j H^{\rm D}_{ij} |\Omega \rangle$
is negative in regions of intense enstrophy. 
The contribution of the isotropic component
is also easy to understand, since 
$\langle \omega_i \omega_j H^{\rm I}_{ij} |\Omega \rangle
= \langle \Omega (\Omega - \Sigma) |\Omega \rangle/6
 = (\Omega^2 - \Omega \langle \Sigma |\Omega \rangle)/6$.
Using $\langle \Sigma |\Omega \rangle \sim \Omega^\gamma$
(with $\gamma < 1$) implies
that  $\langle \omega_i \omega_j H^{\rm I}_{ij} |\Omega \rangle$
is positive in regions of intense enstrophy. 
We verify these expectations in Fig.~\ref{fig:lam2}b-c.

Figure~\ref{fig:lam2}b shows the
conditional expectation  
$\langle \omega_i \omega_j H_{ij}^{\rm D} |\Omega \rangle/\Omega^2$, along 
with the individual contributions
in each eigendirection. It can be clearly seen 
that for large enstrophy,
the contributions from both first and
second eigendirections are essentially zero
and the third eigendirection completely
dominates the overall sum
(as explained earlier). 
This again establishes that the nonlocal
portion of the pressure Hessian actually
enables vortex stretching.
In Fig.~\ref{fig:lam2}c, we compare the 
contributions from the deviatoric and isotropic
components, together with the overall average. 
We notice that the isotropic contribution
is strongly positive, canceling the deviatoric
contribution to give a weakly net positive
average. Once again, this shows that the 
depletion of vortex stretching by pressure Hessian
is local.

It is worth noting that many qualitative aspects of
the previously discussed results 
can be explained by noting
that intense enstrophy is found in tube-like vortices
\cite{Jimenez93, Ishihara07, BPBY2019}, 
for which the Burgers vortex model is a good 
first order approximation \cite{Burgers48, Jimenez93}. 
For the simple case of a Burgers vortex,  
pressure is minimum and constant along the axis
of the vortex. This implies that the
smallest (third) eigenvalue of pressure Hessian is zero
and the corresponding eigenvector is perfectly aligned
with vorticity; whereas the first two eigenvalues
are positive and the corresponding eigenvectors
are perfectly orthogonal to vorticity \cite{andreotti1997}.
Indeed, these expectations are qualitatively consistent with 
the results shown
in Figs.~\ref{fig:align}-\ref{fig:lam1}. 
However, we note that the precise structure of vortices
in turbulence 
is different than Burgers vortex in some very crucial aspects.
For instance, due to the structural properties of 
Burgers vortex, the term $\omega_i \omega_j H_{ij}$
is essentially zero, which is not the case in 
turbulence. 
Additionally, vorticity is perfectly axial in Burgers vortex,
but not real turbulent vortices, 
and there is some noticeable degree of Beltramization 
\cite{Choi:09, BPB2020} -- an effect that is essential
to the self-attenuation mechanism analyzed in \cite{BPB2020, BP2021}. 
In fact, we will discuss in the next subsection that the net positive contribution
from the pressure Hessian as observed in Fig.~\ref{fig:lam2}
is in fact connected to the self-attenuation mechanism 
observed in \cite{BPB2020, BP2021}.

\subsection{Contrasting nonlinear and pressure Hessian contributions}

\begin{figure}
\centering
\includegraphics[width=0.485\linewidth]{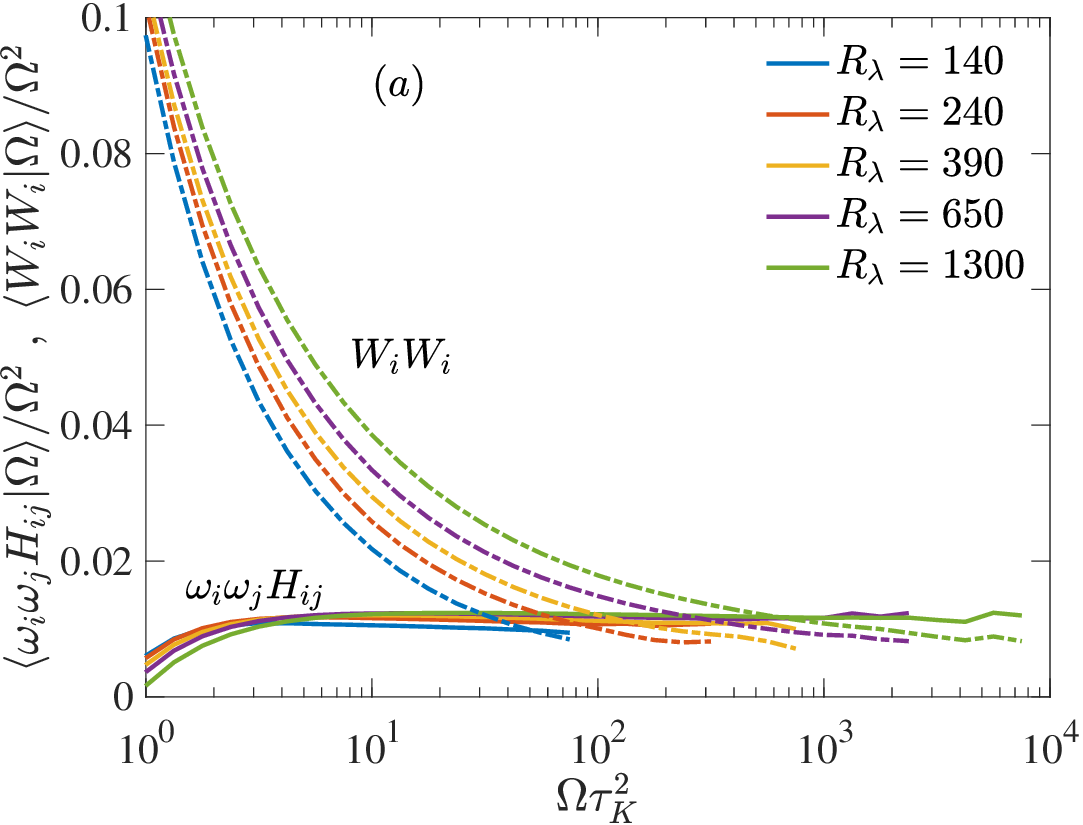} \ \ \
\includegraphics[width=0.485\linewidth]{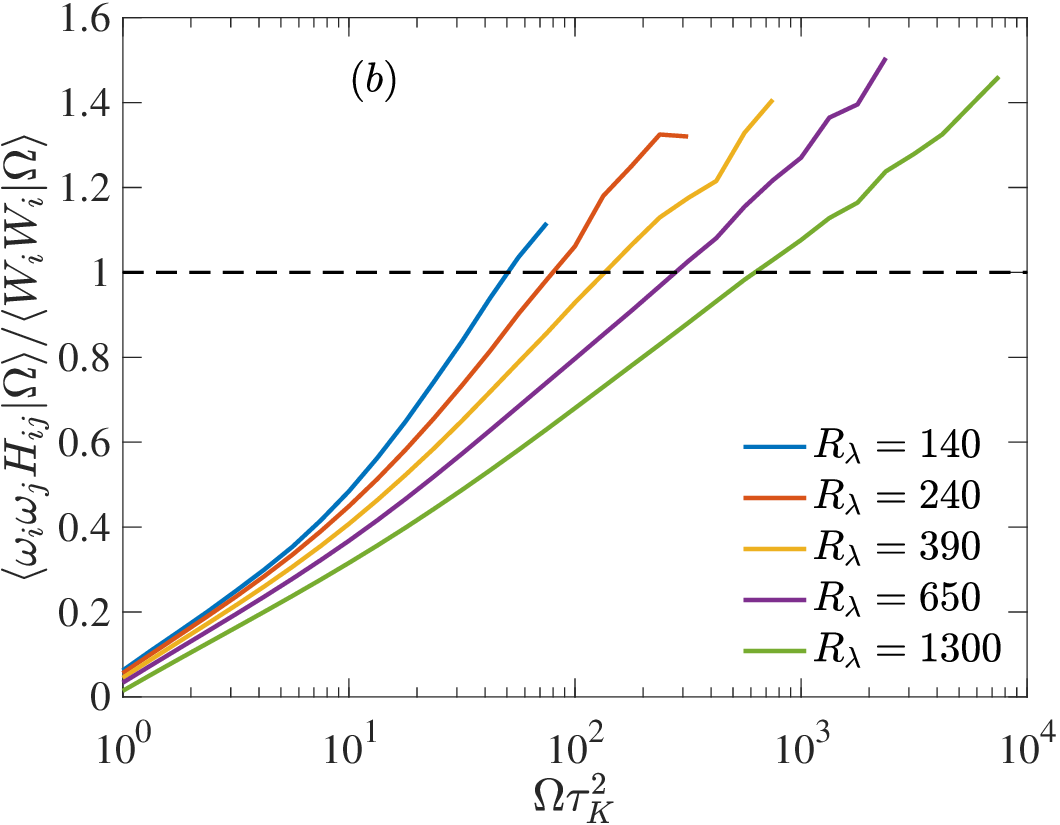} 
\caption{(a) Conditional expectations (given enstrophy $\Omega$)
of the nonlinear and pressure Hessian contributions to 
the dynamics of vortex stretching vector, as given in Eq.~\eqref{eq:w2}.
(a) The ratio of two terms, revealing that pressure Hessian 
contribution overtakes the nonlinear contribution at large $\Omega$.
}
\label{fig:ka}
\end{figure}

Table~\ref{tab:un} demonstrates that 
the (unconditional) contribution $\langle W_i W_i \rangle$
is substantially larger than $\langle \omega_i \omega_j H_{ij}\rangle $
at all $\re$.  A comparison of the conditional result
is first made in Fig.~\ref{fig:ka}a.
Evidently, the nonlinear term dominates the pressure Hessian 
term at $\Omega \tau_K^2 \simeq 1$, in essential agreement
with the observation in
Table~\ref{tab:un}. However, 
for intense enstrophy, we notice a substantial
reduction in the nonlinear term  $\langle W_i W_i | \Omega \rangle/\Omega^2$ relative
to the pressure Hessian term  $\langle \omega_i \omega_j H_{ij} | \Omega \rangle/\Omega^2$,
with the latter being approximately constant.
Since $W_i = \omega_j S_{ij}$,
it follows that 
$\langle W_i W_i |\Omega \rangle \sim 
\langle \Omega \Sigma |\Omega \rangle \sim \Omega^{1+\gamma}$.
In contrast, we find $\langle \omega_i \omega_j H_{ij} |\Omega \rangle \sim \Omega^2$.
Thus, the observation in Fig.~\ref{fig:ka}a  essentially 
shows that the pressure Hessian term 
grows significantly faster than the nonlinear term as $\Omega$ increases,
such that for large enough $\Omega$, the pressure Hessian 
contribution eventually becomes stronger than the nonlinear
contribution (as indeed seen in Fig.~\ref{fig:ka}a).
To show this more clearly, 
Fig.~\ref{fig:ka}b shows the ratio between the two terms,
which is close to zero at $\Omega \tau_K^2 =1$, 
but steadily increases and becomes greater than unity
at large $\Omega$ for every $\re$. 
Hence, the attenuating effect of pressure Hessian
eventually prevails over the nonlinear term
$W_i W_i$. 
Interestingly, the crossover point in $\Omega$
increases with $\re$, which essentially is a reflection
of intermittency and 
$\gamma$ slowly increasing with $\re$.

Further insight on the attenuation induced by the pressure Hessian 
in regions of intense vorticity can be obtained by
rewriting Eq.~\eqref{eq:w2} for the conditional field
\begin{align}
\frac{D \langle \omega_i W_i |\Omega \rangle }{Dt}
=  \langle W_i W_i | \Omega \rangle  
 + (-\langle \omega_i \omega_j H_{ij}^{\rm D} |\Omega \rangle)
 - (\langle \omega_i \omega_j H_{ij}^{\rm I} |\Omega \rangle)
  + \text{viscous terms}
\label{eq:dww}
\end{align}
Note that the l.h.s. is not zero
even in stationary turbulence. 
We essentially observe that the first two terms
on the r.h.s. are positive,
whereas the third term is negative, i.e.,
vortex stretching is enabled by 
the nonlinear term (which is local)
and the deviatoric pressure Hessian (which is nonlocal),
whereas the isotropic pressure Hessian (which is local)
strongly opposes it. 
For weak or moderate $\Omega$, the positive contribution
prevails, resulting in net positive rate of change of
vortex stretching leading to increased vorticity amplification.
However, for large $\Omega$, the negative contribution
from $H_{ij}^{\rm I}$ prevails, leading to negative 
rate of change of vortex stretching. 
In all cases, the viscous terms are ignored, which are
negligibly small at large $\re$ (although not shown, this can
be anticipated).

We stress that, although the pressure Hessian is known to attenuate vortex stretching~\cite{tsi99}, the results in Fig.~\ref{fig:ka} indicate that 
this attenuation overwhelms even the nonlinear terms in regions of most 
intense vorticity. This
points to an inviscid regularizing mechanism, which can be traced 
back to the 
prevalence of the contribution of $H^I$, which can be understood 
as local \cite{Ohkitani:95}.  
A very similar observation for attenuation of vorticity
amplification was also recently uncovered 
in \cite{BPB2020, BP2021}. In these works, the nonlocality
of vortex stretching was analyzed by writing
strain as the Biot-Savart integral of vorticity and
decomposing it into local and nonlocal contributions. The local contribution
is obtained by integrating in a sphere of radius $R$, 
whereas the remaining integral is the nonlocal contribution. 
Thereafter, it was observed that 
vortex stretching is engendered
by the nonlocal contribution and remarkably,
the local contribution acts to attenuate intense vorticity,
also representing an inviscid mechanism to counter vorticity
amplification. 
It stands to reason that the self-attenuation mechanism
identified in  \cite{BPB2020, BP2021} 
is essentially related to the local pressure mechanism
identified in this work. 
An important underlying connection between them is that
they both act only when enstrophy becomes sufficiently
strong and this critical value increases with $\re$ \cite{BPB2020}.
Nevertheless, we note that precisely underpinning 
the causality between the two mechanisms requires
further analysis, particularly by considering
Lagrangian particle trajectories, as evident from 
Eq.~\eqref{eq:dww}. Such an analysis will be considered
in a future work.

\section{Role of pressure Hessian on strain amplification}
\label{sec:strain_hess}

While the previous section focused on the role of the pressure Hessian 
on vorticity amplification, we characterize here the role
of pressure Hessian on 
strain amplification, based on Eq.~\eqref{eq:OS_str}. 
Homogeneity implies that 
$\langle S_{ij} H_{ij} \rangle = 0$, so there is no net
contribution from pressure
Hessian to the budget of $\Sigma$. Nevertheless, the 
situation is quite different when isolating
extreme events, with prior studies
showing that pressure Hessian opposes
strain amplification in regions of intense strain
\cite{nomura:1998, tsi99, Lawson2015, BPB2022}
(and thus, amplifies weak strain).
In our recent work \cite{BPB2022}, 
we already analyzed many aspects of strain
amplification especially by focusing on
individual eigenvalues
of strain.  Here, we present
a complementary
analysis focusing on eigenvalues of pressure Hessian, in the spirit of
the analysis in the previous section.

\subsection{Unconditional statistics}
\label{subsec:u_stat_strain}

\begin{table}
\centering
\begin{tabular}{l|c|c|c}
    $\re$ & $\langle (\mathbf{e}_1^{\rm P} \cdot \mathbf{e}_j)^2 \rangle$ & $\langle (\mathbf{e}_2^{\rm P} \cdot \mathbf{e}_j)^2 \rangle$ &  $\langle (\mathbf{e}_3^{\rm P} \cdot \mathbf{e}_j)^2 \rangle$  \\
    \hline    
140   & 0.231 : 0.401 : 0.368 &  0.377 : 0.373: 0.250 & 0.392 : 0.226 : 0.382 \\
240   & 0.231 : 0.400 : 0.369 &  0.376 : 0.376: 0.249 & 0.393 : 0.224 : 0.383 \\
390   & 0.231 : 0.400 : 0.369 &  0.376 : 0.376: 0.249 & 0.393 : 0.224 : 0.383 \\
650   & 0.231 : 0.400 : 0.369 &  0.375 : 0.377: 0.248 & 0.394 : 0.223 : 0.383 \\
1300  & 0.231 : 0.400 : 0.369 &  0.375 : 0.377: 0.248 & 0.394 : 0.223 : 0.383 \\
\end{tabular}
\caption{
Second moment of alignment cosines between the eigenvectors
of pressure Hessian ($\mathbf{e}_i^{\rm P}$)
and strain ($\mathbf{e}_j$), at various $\re$.
}
\label{tab:un2}
\end{table}

We first analyze the alignments cosines
between eigenvectors of strain and pressure Hessian,
as measured by $\langle (\mathbf{e}_i^{\rm P} \cdot \mathbf{e}_j)^2 \rangle$
for $i,j=1,2,3$. The (unconditional) average of eigenvalues of the 
pressure Hessian can be found in Table~\ref{tab:un}, whereas 
those of strain tensor were previously discussed in \cite{BBP2020}. 
Table~\ref{tab:un2} lists all the nine individual terms
for various $\re$, revealing no particularly
strong alignment between the eigenvectors of strain and pressure
Hessian. The strongest alignment corresponds
to  $\langle (\mathbf{e}_1^{\rm P} \cdot \mathbf{e}_2)^2 \rangle \approx 0.4$,
which is only marginally larger than $1/3$. 
Moreover, all alignment results are virtually independent of
$\re$ as it was earlier the case 
for the alignment between
vorticity and eigenvectors of pressure Hessian.

\begin{figure}
\centering
\includegraphics[width=0.95\linewidth]{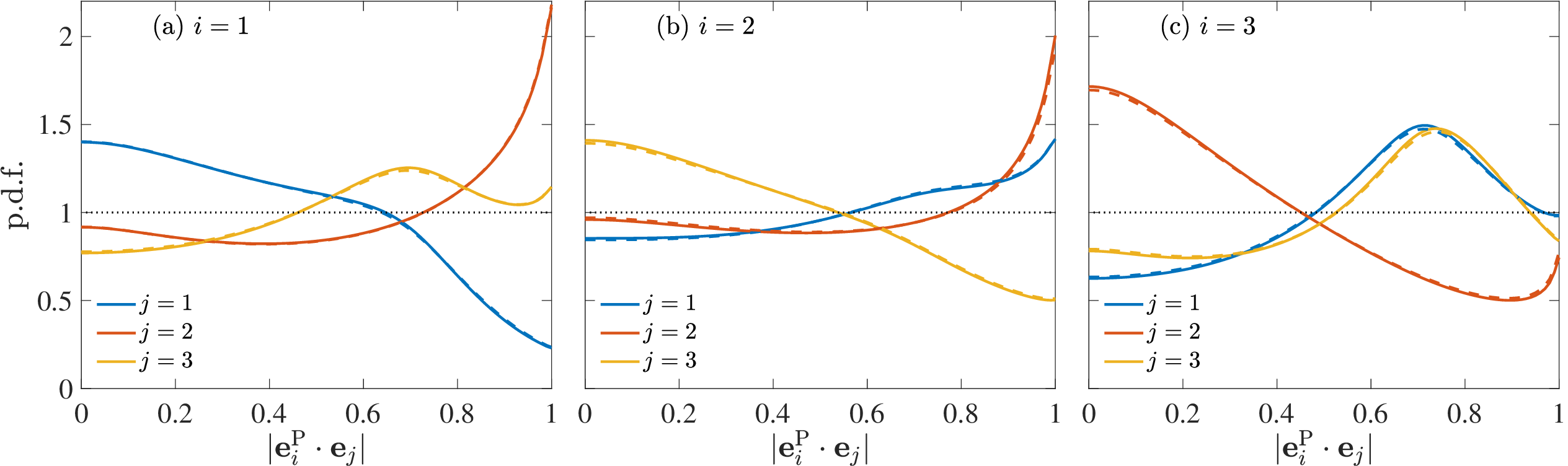}
\caption{
Probability density function (p.d.f.) of
alignment cosines between eigenvectors
of pressure Hessian ($\mathbf{e}_i^{\rm P}$)
and strain ($\mathbf{e}_j$), 
at $\re=1300$ (solid lines) and $\re=140$ (dashed lines). 
}
\label{fig:align_sp}
\end{figure}

To rule out any anomalous behavior, 
Fig.~\ref{fig:align_sp} shows the 
PDFs of the alignment cosines. 
While the distributions are not exactly uniform, it can be seen that
they are essentially consistent with the behavior anticipated from
their second order moments in Table~\ref{tab:un2}, i.e., 
demonstrating some weak preferential alignment 
for moments larger than $1/3$ (and vice versa). 
In all panels, the solid and dashed lines at $\re=1300$ and $140$, respectively,
near perfectly coincide, showing that the alignment results are independent of $\re$.

\subsection{Conditional statistics}
\label{subsec:cond_stat_strain}

\begin{figure}
\centering
\includegraphics[width=0.95\linewidth]{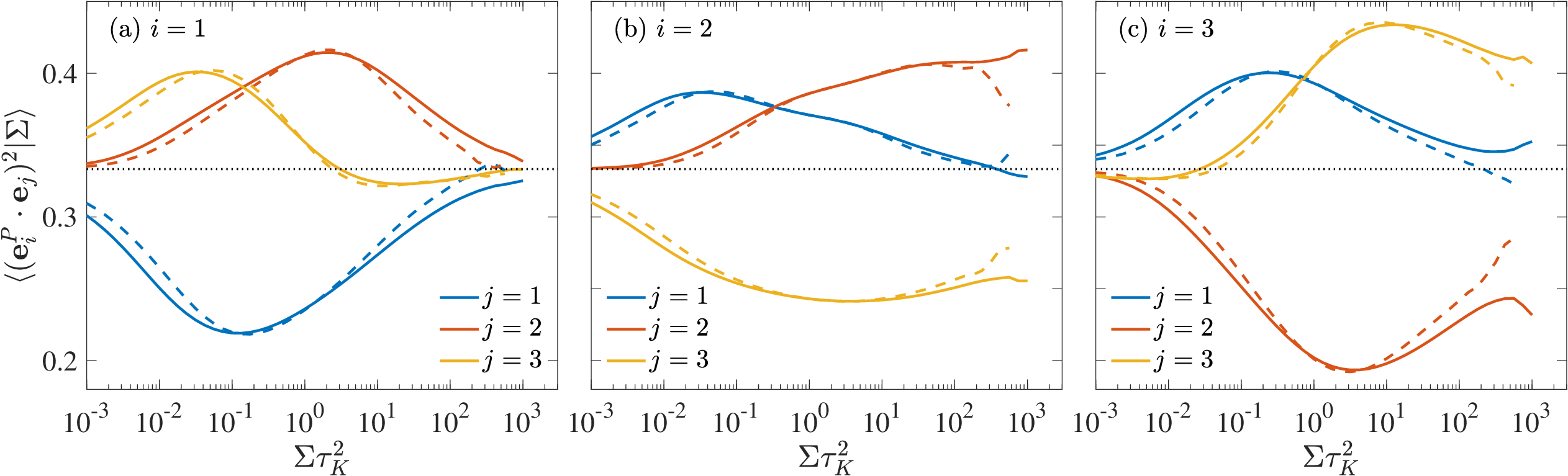}
\caption{
Conditional expectation (given $\Sigma$) 
of second moment of alignment cosines  
 between eigenvectors
of pressure Hessian ($\mathbf{e}_i^{\rm P}$)
and strain ($\mathbf{e}_j$), 
at $\re=1300$ (solid lines) and $\re=650$ (dashed lines). 
}
\label{fig:sh}
\end{figure}

To analyze the extreme strain events, we now consider various
statistics conditioned on 
 $\Sigma \tau_K^2$ (which equals $\Sigma /\langle \Sigma \rangle$).
Figure~\ref{fig:sh} shows the conditional alignment
between eigenvectors of strain and pressure Hessian
as measured by 
$\langle (\mathbf{e}_i^{\rm P} \cdot \mathbf{e}_j)^2 | \Sigma \rangle$. 
For clarity, we only show $\re=1300$ (solid lines) and $\re=650$ (dashed lines).
The dependence on $\re$ is very weak, and results at lower
$\re$ (not shown) essentially follow same trends. 
Overall, the alignment results indicate that
there is no strong preferential alignment between
strain and pressure Hessian, even when extreme events are considered.
The strongest alignments, parallel and orthogonal,
are respectively observed for $j=3$ and $j=2$,
both with $i=3$, but the deviations from $1/3$ remain weak.

\begin{figure}
\centering
\includegraphics[width=0.45\linewidth]{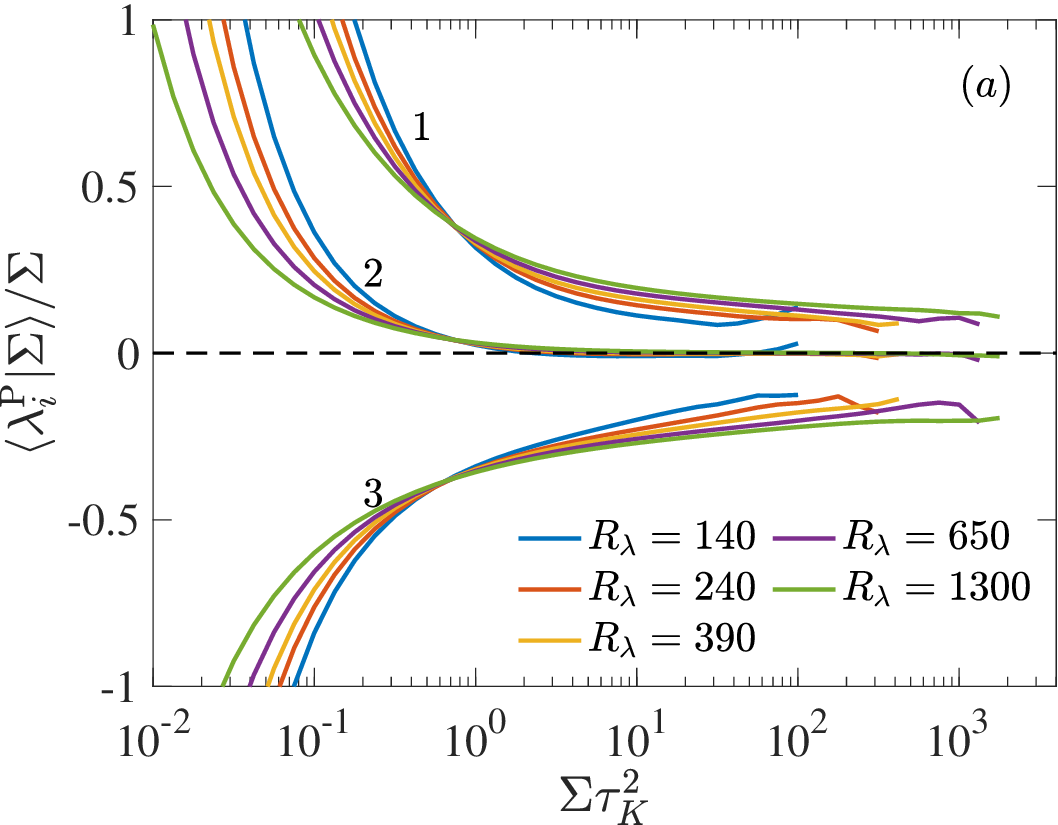} \ \ \ \
\includegraphics[width=0.45\linewidth]{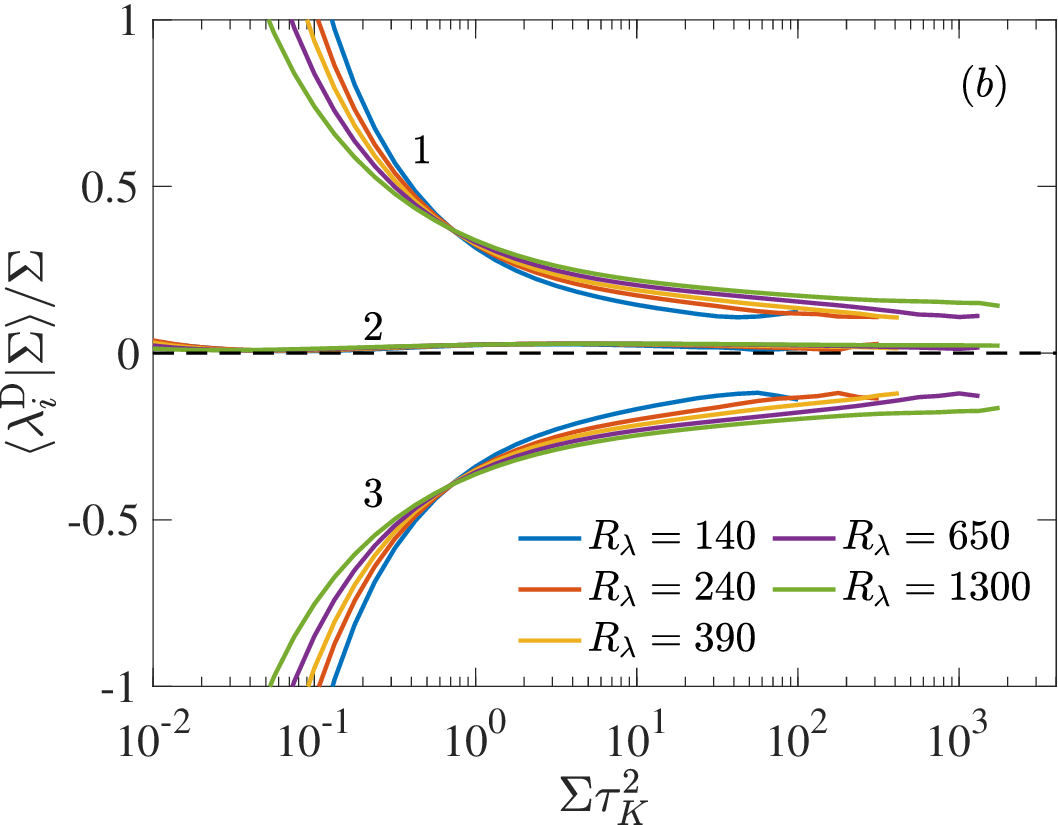}
\caption{
(a) Conditional expectation (given $\Sigma$) of the eigenvalues 
of pressure Hessian, at various $\re$.
(b) Conditional expectation of the eigenvalues of the deviatoric
part.
}
\label{fig:lam_d}
\end{figure}

The conditional expectations of the eigenvalues of the pressure Hessian, 
and of its deviatoric part, are shown in
Figure~\ref{fig:lam_d}a and b, respectively.
In both cases, we observe that the first and third eigenvalues
are strongly positive and negative respectively, and the second
eigenvalue is very close to zero. 
The eigenvalues of pressure Hessian
satisfy $ \langle (\lambda_1^{\rm P} + \lambda_2^{\rm P} + 
\lambda_3^{\rm P}) | \Sigma \rangle 
= \langle (\Omega - \Sigma) | \Sigma \rangle $.
Since $\langle \Omega |\Sigma \rangle \sim \Sigma$, but with a 
prefactor which is slightly smaller than unity \cite{BPB2022, BP2022},
it follows that the sum of eigenvalues $\lambda_i^P$ divided by $\Sigma$ 
is a small, negative constant. Indeed,
this is the observation in Fig.~\ref{fig:lam_d}a,
which shows that 
$\langle -\lambda_3^{\rm P} |\Sigma  \rangle \gtrsim \langle \lambda_1^{\rm P} |\Sigma  \rangle$, whereas 
$\langle \lambda_2^{\rm P} |\Sigma  \rangle \approx 0$.
On the contrary, 
the sum of eigenvalues of the deviatoric part is exactly zero.
Indeed, Fig.~\ref{fig:lam_d}b conforms with this expectation, 
with
$\langle -\lambda_3^{\rm D} |\Sigma  \rangle \gtrsim 
\langle \lambda_1^{\rm D} |\Sigma  \rangle$ being still true, but 
$\langle \lambda_2^{\rm D} |\Sigma  \rangle $ is
weakly positive, ensuring that the sum of the eigenvalues is zero.

\begin{figure}
\centering
\includegraphics[width=0.485\linewidth]{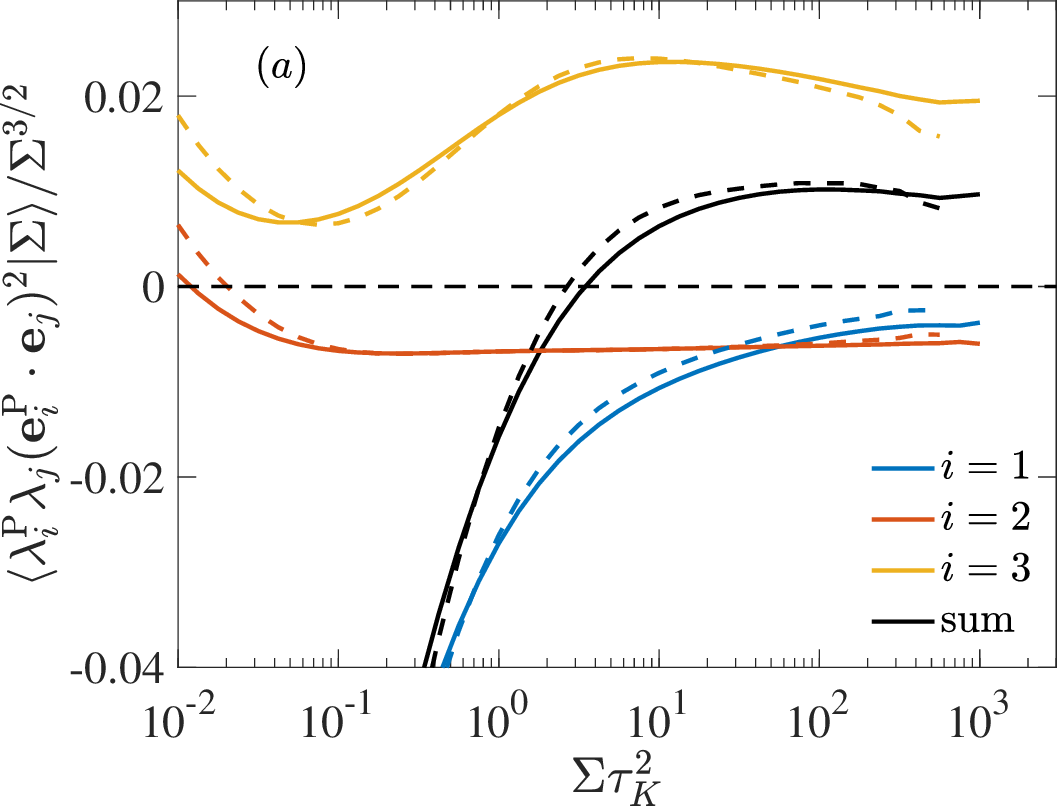} \ \ \ 
\includegraphics[width=0.485\linewidth]{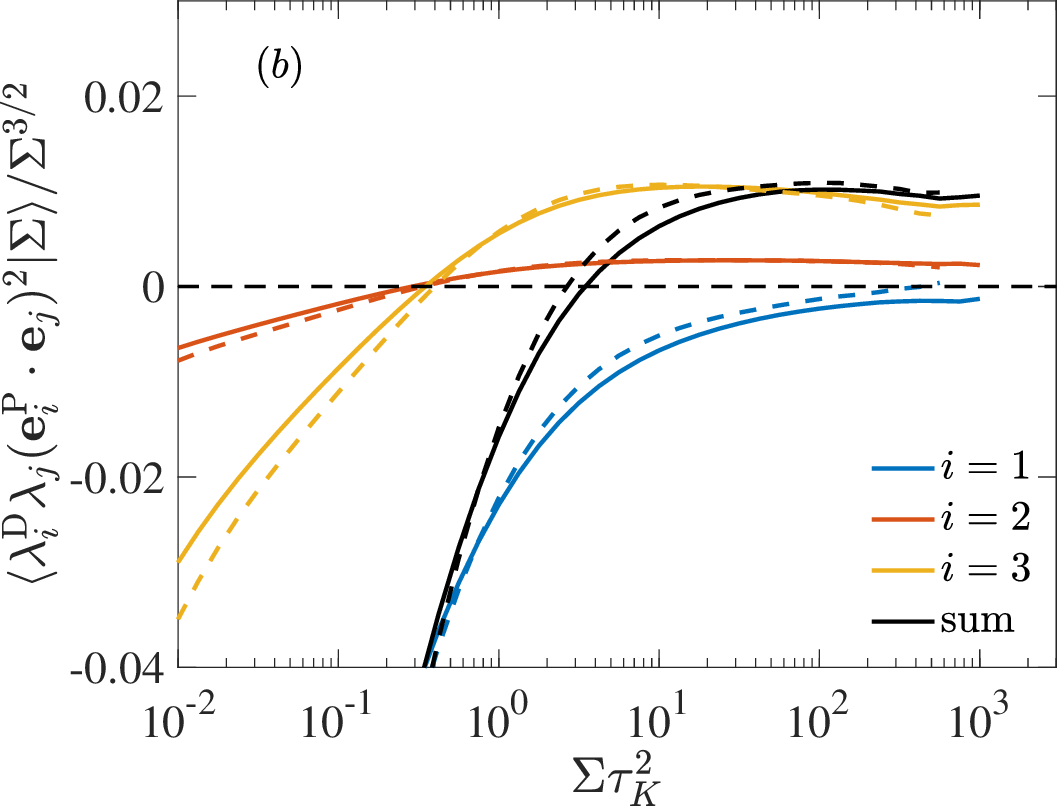}
\caption{
Individual contributions to 
(a) $\langle S_{ij} H_{ij} |\Sigma \rangle  =  
\langle \lambda_i^{\rm P} \lambda_j ( \mathbf{e}_i^{\rm P} \cdot \mathbf{e}_j )^2 | \Sigma \rangle$,  and 
(b) $\langle S_{ij} H_{ij}^{\rm D} |\Sigma \rangle  =  
\langle \lambda_i^{\rm D} \lambda_j ( \mathbf{e}_i^{\rm D} \cdot \mathbf{e}_j )^2 | \Sigma \rangle$, from their eigendirections. 
Note, the summation is only implied over $j$, and results
for each $i$ are shown. 
The quantities have been rescaled by $\Sigma^{3/2}$.
The solid lines are for $\re=1300$
and dashed lines for $\re=650$. 
The black lines given the total sum over all eigendirections.
}
\label{fig:sh_d}
\end{figure}

Figure~\ref{fig:sh_d} shows conditional expectation of the correlation
$S_{ij} H_{ij}$, together with the individual contributions
from the eigendirections of pressure Hessian; panel a  shows the
result for pressure Hessian and its eigenvalues and 
panel b shows the corresponding result for the deviatoric part. 
Note that despite $S_{ij} H_{ij} = S_{ij} H_{ij}^{\rm D}$,
the individual contributions from their respective eigendirections
differ. 
Given the lack of any strong alignment between 
strain and pressure Hessian eigenvectors, it can be anticipated
that the largest contribution to $S_{ij} H_{ij}$ 
would arise from their largest eigenvalues, i.e., 
the product $\lambda_3^{\rm P} \lambda_3$ 
(or $\lambda_3^{\rm D} \lambda_3$). Additionally, this 
contribution would be positive, since both these
eigenvalues are negative. 
Indeed, Fig.~\ref{fig:sh_d}a-b confirms this expectation.
For both panels, the largest contribution is positive
and corresponds to $i=3$. 
In contrast, the contribution for $i=1$ is negative, 
since it is dominated by the product $\lambda_1^{\rm P} \lambda_3$,
(or $\lambda_1^{\rm D} \lambda_3$ for panel b). 
The behavior for $i=2$ is not as straightforward to predict,
since the alignments are nontrivial for this case.
Interestingly, we observe that the contribution
from $i=2$ is negative in panel a, 
but weakly positive in panel b. 
For large $\Sigma$, all the contributions
(as divided by $\Sigma^{3/2}$) appear
approximately constant, implying a simple
scaling \cite{BPB2022}.

\begin{figure}
\centering
\includegraphics[width=0.55\linewidth]{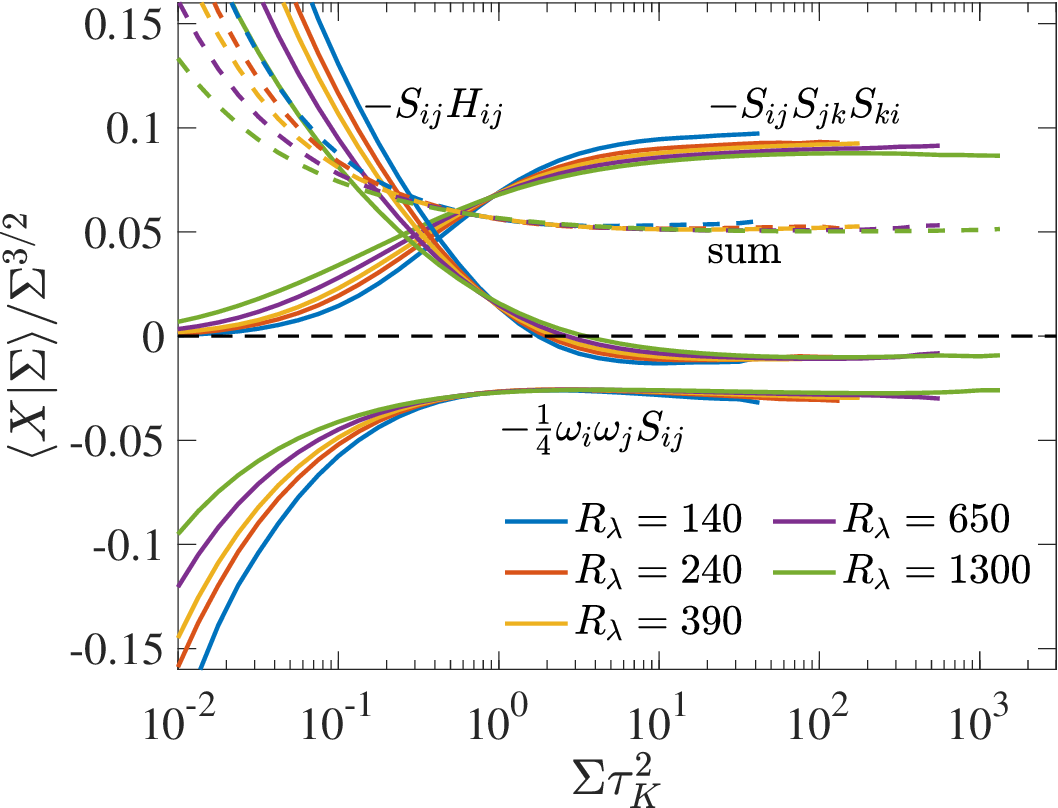} 
\caption{Conditional expectations (given $\Sigma$) of various
inviscid terms on the r.h.s. of Eq.~\eqref{eq:OS_str}. All quantities
are normalized by $\Sigma^{3/2}$ revealing a plateau like behavior
for $\Sigma \tau_K^2 > 1$.
}
\label{fig:bud_d}
\end{figure}

The main observation from Fig.~\ref{fig:sh_d}
is that the pressure Hessian term attenuates strong strain
and amplifies weak strain
(since the net effect has to be zero). 
Nevertheless, as evident from Eq.~\eqref{eq:OS_str}, 
strain amplification involves other nonlinear mechanisms.
Their relative amplitudes were already compared in \cite{BPB2022}.
For completeness, 
Fig.~\ref{fig:bud_d}
shows the conditional expectations of 
the various inviscid terms involved
in the budget equation for strain
(together with their corresponding signs). All terms
are once again normalized by $\Sigma^{3/2}$. 
It can be seen that the self-amplification 
term is the largest contribution and the primary
mechanism for generating intense strain. 
In contrast, the vortex stretching 
contribution always opposes strain amplification,
whereas the pressure Hessian term opposes
intense strain but amplifies weak strain. 
When considering strong strain events
($\Sigma \tau_K^2 > 1$), the pressure Hessian 
term is noticeably weaker than vortex stretching
term, and both are considerably weaker than 
the strain self-amplification term.

\section{Summary and conclusions}

The pressure field, via its Hessian tensor, 
plays a direct role in the dynamics
of velocity gradient tensor $A_{ij}$ in turbulence. 
Since pressure Hessian $H_{ij}$ is a symmetric tensor,
its effect on the amplification of strain $S_{ij}$ (the symmetric
part of $A_{ij}$) is explicit. However,
its influence on amplification of vorticity vector $\omega_i$ 
(the skew-symmetric part of $A_{ij}$) 
comes as a second order effect through the dynamics of the vortex
stretching vector $W_i = \omega_j S_{ij}$.
However, 
studies directly investigating the role of pressure Hessian
on gradient amplification 
have been limited in the literature and when available, 
restricted to low Reynolds numbers.  In this paper, 
utilizing a massive DNS
database of isotropic turbulence across a wide
range of Taylor-scale Reynolds numbers 
($140 \le \re \le 1300$), 
we systematically investigate various statistical
correlations underpinning the role of pressure Hessian
on gradient amplification and formation of extreme events.

Overall, pressure Hessian acts to deplete vortex
stretching (and also strain amplification, which is discussed soon). 
Decomposing the pressure Hessian into its isotropic (local) 
and deviatoric (nonlocal) components 
reveals that the former opposes vortex stretching, whereas 
the latter favors it.
There is strong cancellation between the two, such that the isotropic
contribution ultimately dominates, although the net 
effect is significantly weaker than 
the nonlinear mechanism (which always enables vortex stretching).
However, when the statistics are conditioned
on enstrophy $\Omega =\omega_i \omega_i$, we find that
the inhibiting effect of pressure Hessian 
becomes dominant over the nonlinear mechanism, essentially
leading to net depletion of vortex stretching.
This depletion, which comes from local isotropic
part, suggests a natural connection to the self-attenuation
mechanism recently identified in
\cite{BPB2020, BP2021}
whereby it was shown that the self-induced local strain
in regions of intense vorticity acts to attenuate further
growth of vorticity.
However, additional investigation is required to 
reinforce this connection, particularly by analyzing the dynamics
of velocity gradients in the 
Lagrangian framework, which will be the subject of future work.

We further analyze the contribution of pressure Hessian to vortex stretching 
in its eigenframe. For the mean field, 
the sum of eigenvalues is zero, resulting in first (largest) and third
(smallest) eigenvalues being nearly equal in magnitude, but opposite in sign;
the second (intermediate) eigenvalue is weakly positive but essentially close to zero. 
No particular strong alignment is observed between vorticity
and eigenvectors of pressure Hessian. However, when considering
conditional statistics, the third eigenvalue becomes close to zero,
and both first and second eigenvalues are positive. At the same time,
vorticity near perfectly aligns with the third eigenvector (while being
orthogonal to other two). Structurally, this conforms
with the well known notion that intense vorticity in turbulence
is arranged in tubes, which approximately correspond to Burgers vortex.
However, key differences remain, which essentially give rise to the 
self-attenuation mechanism, as elucidated by \cite{BPB2020}. 
Considering the deviatoric and isotropic components
separately allows for a simpler interpretation of how pressure Hessian
overall affects vortex stretching.
Overall, not particularly strong dependence on Reynolds number
is observed for all the statistics considered.

We also investigate the role of pressure Hessian on strain
amplification, which is more direct, since both tensor are symmetric. 
Overall, pressure Hessian
does not contribute to the budget of strain, as their correlation
is zero (using homogeneity). However, when considering conditional statistics,
we find that pressure Hessian acts to oppose intense strain,
but amplifies weak strain, essentially driving strain fluctuations
towards the mean amplitude. 
Complementary to the analysis presented in~\cite{BPB2022}, 
we focus here on the eigenframe of pressure
Hessian. As before, the first and third eigenvalues of pressure Hessian
are comparable in magnitude, but opposite in sign. This behavior
also applies when conditioning on strain magnitude.
No strong alignments are observed between eigenvectors
of strain and pressure Hessian, both for unconditional and conditional
fields. Because of this, the resulting behavior 
strain and pressure Hessian correlation can be deduced
simply from the product of their eigenvalues. 
We contrast the strength of pressure Hessian term
with other nonlinear mechanisms controlling strain amplification,
viz. the self-amplification term and vortex stretching. 
Evidently, the self-amplification term always amplifies
strain; however, vortex stretching always opposes it,
with its magnitude significantly weaker. The opposition from
pressure Hessian term is even weaker.

Finally, we note that many of the conditional statistics
investigated here follow simple power law dependencies
on vorticity or strain magnitude (when considering
extreme events). We have identified these power laws
in various figures as necessary. 
These results should prove valuable in statistical modeling
of enstrophy or energy dissipation rate, particularly
in PDF methods \cite{popebook}. 
Additionally, the various conditional statistics of pressure Hessian
should also provide valuable benchmarks for Lagrangian
modeling of velocity gradient dynamics \cite{Meneveau11}.
These aspects will be explored in future work.

\section*{Acknowledgements}
We gratefully acknowledge the Gauss Centre for Supercomputing e.V.
for providing computing time on the supercomputers JUWELS and JUQUEEN
at Jülich Supercomputing Centre (JSC), where the simulations and data
analyses reported in this paper were performed.


\begin{thebibliography}{50}%
\makeatletter
\providecommand \@ifxundefined [1]{%
 \@ifx{#1\undefined}
}%
\providecommand \@ifnum [1]{%
 \ifnum #1\expandafter \@firstoftwo
 \else \expandafter \@secondoftwo
 \fi
}%
\providecommand \@ifx [1]{%
 \ifx #1\expandafter \@firstoftwo
 \else \expandafter \@secondoftwo
 \fi
}%
\providecommand \natexlab [1]{#1}%
\providecommand \enquote  [1]{``#1''}%
\providecommand \bibnamefont  [1]{#1}%
\providecommand \bibfnamefont [1]{#1}%
\providecommand \citenamefont [1]{#1}%
\providecommand \href@noop [0]{\@secondoftwo}%
\providecommand \href [0]{\begingroup \@sanitize@url \@href}%
\providecommand \@href[1]{\@@startlink{#1}\@@href}%
\providecommand \@@href[1]{\endgroup#1\@@endlink}%
\providecommand \@sanitize@url [0]{\catcode `\\12\catcode `\$12\catcode
  `\&12\catcode `\#12\catcode `\^12\catcode `\_12\catcode `\%12\relax}%
\providecommand \@@startlink[1]{}%
\providecommand \@@endlink[0]{}%
\providecommand \url  [0]{\begingroup\@sanitize@url \@url }%
\providecommand \@url [1]{\endgroup\@href {#1}{\urlprefix }}%
\providecommand \urlprefix  [0]{URL }%
\providecommand \Eprint [0]{\href }%
\providecommand \doibase [0]{https://doi.org/}%
\providecommand \selectlanguage [0]{\@gobble}%
\providecommand \bibinfo  [0]{\@secondoftwo}%
\providecommand \bibfield  [0]{\@secondoftwo}%
\providecommand \translation [1]{[#1]}%
\providecommand \BibitemOpen [0]{}%
\providecommand \bibitemStop [0]{}%
\providecommand \bibitemNoStop [0]{.\EOS\space}%
\providecommand \EOS [0]{\spacefactor3000\relax}%
\providecommand \BibitemShut  [1]{\csname bibitem#1\endcsname}%
\let\auto@bib@innerbib\@empty
\bibitem [{\citenamefont {Frisch}(1995)}]{Frisch95}%
  \BibitemOpen
  \bibfield  {author} {\bibinfo {author} {\bibfnamefont {U.}~\bibnamefont
  {Frisch}},\ }\href@noop {} {\emph {\bibinfo {title} {Turbulence: the legacy
  of {Kolmogorov}}}}\ (\bibinfo  {publisher} {Cambridge University Press},\
  \bibinfo {address} {Cambridge},\ \bibinfo {year} {1995})\BibitemShut
  {NoStop}%
\bibitem [{\citenamefont {Sreenivasan}\ and\ \citenamefont
  {Antonia}(1997)}]{Sreeni97}%
  \BibitemOpen
  \bibfield  {author} {\bibinfo {author} {\bibfnamefont {K.~S.}\ \bibnamefont
  {Sreenivasan}}\ and\ \bibinfo {author} {\bibfnamefont {R.~A.}\ \bibnamefont
  {Antonia}},\ }\bibfield  {title} {\bibinfo {title} {The phenomenology of
  small-scale turbulence},\ }\href@noop {} {\bibfield  {journal} {\bibinfo
  {journal} {Annu.~Rev.~Fluid~Mech.}\ }\textbf {\bibinfo {volume} {29}},\
  \bibinfo {pages} {435} (\bibinfo {year} {1997})}\BibitemShut {NoStop}%
\bibitem [{\citenamefont {Falkovich}\ \emph {et~al.}(2001)\citenamefont
  {Falkovich}, \citenamefont {Gaw\ifmmode~\mbox{\c{e}}\else \c{e}\fi{}dzki},\
  and\ \citenamefont {Vergassola}}]{falkovich01}%
  \BibitemOpen
  \bibfield  {author} {\bibinfo {author} {\bibfnamefont {G.}~\bibnamefont
  {Falkovich}}, \bibinfo {author} {\bibfnamefont {K.}~\bibnamefont
  {Gaw\ifmmode~\mbox{\c{e}}\else \c{e}\fi{}dzki}},\ and\ \bibinfo {author}
  {\bibfnamefont {M.}~\bibnamefont {Vergassola}},\ }\bibfield  {title}
  {\bibinfo {title} {Particles and fields in fluid turbulence},\ }\href@noop {}
  {\bibfield  {journal} {\bibinfo  {journal} {Rev. Mod. Phys.}\ }\textbf
  {\bibinfo {volume} {73}},\ \bibinfo {pages} {913} (\bibinfo {year}
  {2001})}\BibitemShut {NoStop}%
\bibitem [{\citenamefont {Wallace}(2009)}]{Wallace09}%
  \BibitemOpen
  \bibfield  {author} {\bibinfo {author} {\bibfnamefont {J.~M.}\ \bibnamefont
  {Wallace}},\ }\bibfield  {title} {\bibinfo {title} {Twenty years of
  experimental and direct numerical simulation access to the velocity gradient
  tensor: What have we learned about turbulence?},\ }\href@noop {} {\bibfield
  {journal} {\bibinfo  {journal} {Phys. Fluids}\ }\textbf {\bibinfo {volume}
  {21}},\ \bibinfo {pages} {021301} (\bibinfo {year} {2009})}\BibitemShut
  {NoStop}%
\bibitem [{\citenamefont {Tsinober}(2009)}]{Tsi2009}%
  \BibitemOpen
  \bibfield  {author} {\bibinfo {author} {\bibfnamefont {A.}~\bibnamefont
  {Tsinober}},\ }\href@noop {} {\emph {\bibinfo {title} {An Informal Conceptual
  Introduction to Turbulence}}}\ (\bibinfo  {publisher} {Springer},\ \bibinfo
  {address} {Berlin},\ \bibinfo {year} {2009})\BibitemShut {NoStop}%
\bibitem [{\citenamefont {Meneveau}(2011)}]{Meneveau11}%
  \BibitemOpen
  \bibfield  {author} {\bibinfo {author} {\bibfnamefont {C.}~\bibnamefont
  {Meneveau}},\ }\bibfield  {title} {\bibinfo {title} {Lagrangian dynamics and
  models of the velocity gradient tensor in turbulent flows},\ }\href@noop {}
  {\bibfield  {journal} {\bibinfo  {journal} {Annu. Rev. Fluid Mech.}\ }\textbf
  {\bibinfo {volume} {43}},\ \bibinfo {pages} {219} (\bibinfo {year}
  {2011})}\BibitemShut {NoStop}%
\bibitem [{\citenamefont {Falkovich}\ \emph {et~al.}(2002)\citenamefont
  {Falkovich}, \citenamefont {Fouxon},\ and\ \citenamefont
  {Stepanov}}]{falkovich02}%
  \BibitemOpen
  \bibfield  {author} {\bibinfo {author} {\bibfnamefont {G.}~\bibnamefont
  {Falkovich}}, \bibinfo {author} {\bibfnamefont {A.}~\bibnamefont {Fouxon}},\
  and\ \bibinfo {author} {\bibfnamefont {M.~G.}\ \bibnamefont {Stepanov}},\
  }\bibfield  {title} {\bibinfo {title} {Acceleration of rain initiation by
  cloud turbulence},\ }\href@noop {} {\bibfield  {journal} {\bibinfo  {journal}
  {Nature}\ }\textbf {\bibinfo {volume} {419}},\ \bibinfo {pages} {151}
  (\bibinfo {year} {2002})}\BibitemShut {NoStop}%
\bibitem [{\citenamefont {Hamlington}\ \emph {et~al.}(2012)\citenamefont
  {Hamlington}, \citenamefont {Poludnenko},\ and\ \citenamefont
  {Oran}}]{hamlington12}%
  \BibitemOpen
  \bibfield  {author} {\bibinfo {author} {\bibfnamefont {P.~E.}\ \bibnamefont
  {Hamlington}}, \bibinfo {author} {\bibfnamefont {A.~Y.}\ \bibnamefont
  {Poludnenko}},\ and\ \bibinfo {author} {\bibfnamefont {E.~S.}\ \bibnamefont
  {Oran}},\ }\bibfield  {title} {\bibinfo {title} {Intermittency in premixed
  turbulent reacting flows},\ }\href@noop {} {\bibfield  {journal} {\bibinfo
  {journal} {Phys.~Fluids}\ }\textbf {\bibinfo {volume} {24}},\ \bibinfo
  {pages} {075111} (\bibinfo {year} {2012})}\BibitemShut {NoStop}%
\bibitem [{\citenamefont {Buaria}\ \emph {et~al.}(2015)\citenamefont {Buaria},
  \citenamefont {Sawford},\ and\ \citenamefont {Yeung}}]{BSY.2015}%
  \BibitemOpen
  \bibfield  {author} {\bibinfo {author} {\bibfnamefont {D.}~\bibnamefont
  {Buaria}}, \bibinfo {author} {\bibfnamefont {B.~L.}\ \bibnamefont
  {Sawford}},\ and\ \bibinfo {author} {\bibfnamefont {P.~K.}\ \bibnamefont
  {Yeung}},\ }\bibfield  {title} {\bibinfo {title} {Characteristics of backward
  and forward two-particle relative dispersion in turbulence at different
  {R}eynolds numbers},\ }\href@noop {} {\bibfield  {journal} {\bibinfo
  {journal} {Phys. Fluids}\ }\textbf {\bibinfo {volume} {27}},\ \bibinfo
  {pages} {105101} (\bibinfo {year} {2015})}\BibitemShut {NoStop}%
\bibitem [{\citenamefont {Voth}\ and\ \citenamefont
  {Soldati}(2017)}]{voth2017}%
  \BibitemOpen
  \bibfield  {author} {\bibinfo {author} {\bibfnamefont {G.~A.}\ \bibnamefont
  {Voth}}\ and\ \bibinfo {author} {\bibfnamefont {A.}~\bibnamefont {Soldati}},\
  }\bibfield  {title} {\bibinfo {title} {Anisotropic particles in turbulence},\
  }\href@noop {} {\bibfield  {journal} {\bibinfo  {journal} {Annu.~Rev.~Fluid
  Mech.}\ }\textbf {\bibinfo {volume} {49}},\ \bibinfo {pages} {249} (\bibinfo
  {year} {2017})}\BibitemShut {NoStop}%
\bibitem [{\citenamefont {Buaria}\ \emph {et~al.}(2021)\citenamefont {Buaria},
  \citenamefont {Clay}, \citenamefont {Sreenivasan},\ and\ \citenamefont
  {Yeung}}]{BCSY2021a}%
  \BibitemOpen
  \bibfield  {author} {\bibinfo {author} {\bibfnamefont {D.}~\bibnamefont
  {Buaria}}, \bibinfo {author} {\bibfnamefont {M.~P.}\ \bibnamefont {Clay}},
  \bibinfo {author} {\bibfnamefont {K.~R.}\ \bibnamefont {Sreenivasan}},\ and\
  \bibinfo {author} {\bibfnamefont {P.~K.}\ \bibnamefont {Yeung}},\ }\bibfield
  {title} {\bibinfo {title} {Small-scale isotropy and ramp-cliff structures in
  scalar turbulence},\ }\href@noop {} {\bibfield  {journal} {\bibinfo
  {journal} {Phys.~Rev.~Lett.}\ }\textbf {\bibinfo {volume} {126}},\ \bibinfo
  {pages} {034504} (\bibinfo {year} {2021})}\BibitemShut {NoStop}%
\bibitem [{\citenamefont {Kolmogorov}(1941)}]{K41a}%
  \BibitemOpen
  \bibfield  {author} {\bibinfo {author} {\bibfnamefont {A.~N.}\ \bibnamefont
  {Kolmogorov}},\ }\bibfield  {title} {\bibinfo {title} {The local structure of
  turbulence in an incompressible fluid for very large reynolds numbers},\
  }\href@noop {} {\bibfield  {journal} {\bibinfo  {journal} {Dokl. Akad. Nauk.
  SSSR}\ }\textbf {\bibinfo {volume} {30}},\ \bibinfo {pages} {299} (\bibinfo
  {year} {1941})}\BibitemShut {NoStop}%
\bibitem [{\citenamefont {Gibbon}\ \emph {et~al.}(2008)\citenamefont {Gibbon},
  \citenamefont {Bustamante},\ and\ \citenamefont {Kerr}}]{gibbon:2008}%
  \BibitemOpen
  \bibfield  {author} {\bibinfo {author} {\bibfnamefont {J.~D.}\ \bibnamefont
  {Gibbon}}, \bibinfo {author} {\bibfnamefont {M.}~\bibnamefont {Bustamante}},\
  and\ \bibinfo {author} {\bibfnamefont {R.~M.}\ \bibnamefont {Kerr}},\
  }\bibfield  {title} {\bibinfo {title} {The three-dimensional {Euler}
  equations: singular or non-singular?},\ }\href@noop {} {\bibfield  {journal}
  {\bibinfo  {journal} {Nonlinearity}\ }\textbf {\bibinfo {volume} {21}},\
  \bibinfo {pages} {T123} (\bibinfo {year} {2008})}\BibitemShut {NoStop}%
\bibitem [{\citenamefont {Doering}(2009)}]{doering2009}%
  \BibitemOpen
  \bibfield  {author} {\bibinfo {author} {\bibfnamefont {C.~R.}\ \bibnamefont
  {Doering}},\ }\bibfield  {title} {\bibinfo {title} {{The {3D} {Navier-Stokes}
  problem}},\ }\href@noop {} {\bibfield  {journal} {\bibinfo  {journal}
  {Annu.~Rev.~Fluid Mech.}\ }\textbf {\bibinfo {volume} {41}},\ \bibinfo
  {pages} {109} (\bibinfo {year} {2009})}\BibitemShut {NoStop}%
\bibitem [{\citenamefont {Fefferman}(2006)}]{Fefferman}%
  \BibitemOpen
  \bibfield  {author} {\bibinfo {author} {\bibfnamefont {C.}~\bibnamefont
  {Fefferman}},\ }\bibinfo {title} {Existence and smoothness of the
  {Navier-Stokes} equations}\ (\bibinfo  {publisher} {Clay Mathematical
  Institute},\ \bibinfo {address} {Cambridge, MA},\ \bibinfo {year}
  {2006})\BibitemShut {NoStop}%
\bibitem [{\citenamefont {Nomura}\ and\ \citenamefont
  {Post}(1998)}]{nomura:1998}%
  \BibitemOpen
  \bibfield  {author} {\bibinfo {author} {\bibfnamefont {K.~K.}\ \bibnamefont
  {Nomura}}\ and\ \bibinfo {author} {\bibfnamefont {G.~K.}\ \bibnamefont
  {Post}},\ }\bibfield  {title} {\bibinfo {title} {The structure and dynamics
  of vorticity and rate of strain in incompressible homogeneous turbulence},\
  }\href@noop {} {\bibfield  {journal} {\bibinfo  {journal} {J. Fluid Mech.}\
  }\textbf {\bibinfo {volume} {377}},\ \bibinfo {pages} {65} (\bibinfo {year}
  {1998})}\BibitemShut {NoStop}%
\bibitem [{\citenamefont {Kalelkar}(2006)}]{kalelkar2006}%
  \BibitemOpen
  \bibfield  {author} {\bibinfo {author} {\bibfnamefont {C.}~\bibnamefont
  {Kalelkar}},\ }\bibfield  {title} {\bibinfo {title} {Statistics of pressure
  fluctuations in decaying isotropic turbulence},\ }\href@noop {} {\bibfield
  {journal} {\bibinfo  {journal} {Phys.~Rev.~E}\ }\textbf {\bibinfo {volume}
  {73}},\ \bibinfo {pages} {046301} (\bibinfo {year} {2006})}\BibitemShut
  {NoStop}%
\bibitem [{\citenamefont {Lawson}\ and\ \citenamefont
  {Dawson}(2015)}]{Lawson2015}%
  \BibitemOpen
  \bibfield  {author} {\bibinfo {author} {\bibfnamefont {J.~M.}\ \bibnamefont
  {Lawson}}\ and\ \bibinfo {author} {\bibfnamefont {J.~R.}\ \bibnamefont
  {Dawson}},\ }\bibfield  {title} {\bibinfo {title} {On velocity gradient
  dynamics and turbulent structure},\ }\href@noop {} {\bibfield  {journal}
  {\bibinfo  {journal} {J.~Fluid Mech.}\ }\textbf {\bibinfo {volume} {780}},\
  \bibinfo {pages} {60–98} (\bibinfo {year} {2015})}\BibitemShut {NoStop}%
\bibitem [{\citenamefont {Carbone}\ \emph {et~al.}(2020)\citenamefont
  {Carbone}, \citenamefont {Iovieno},\ and\ \citenamefont
  {Bragg}}]{Carbone:20b}%
  \BibitemOpen
  \bibfield  {author} {\bibinfo {author} {\bibfnamefont {M.}~\bibnamefont
  {Carbone}}, \bibinfo {author} {\bibfnamefont {M.}~\bibnamefont {Iovieno}},\
  and\ \bibinfo {author} {\bibfnamefont {A.~D.}\ \bibnamefont {Bragg}},\
  }\bibfield  {title} {\bibinfo {title} {Symmetry transformation and
  dimensionality reduction of the anistropic pressure hessian},\ }\href@noop {}
  {\bibfield  {journal} {\bibinfo  {journal} {J.~Fluid Mech.}\ }\textbf
  {\bibinfo {volume} {900}},\ \bibinfo {pages} {A38} (\bibinfo {year}
  {2020})}\BibitemShut {NoStop}%
\bibitem [{\citenamefont {Buaria}\ \emph {et~al.}(2022)\citenamefont {Buaria},
  \citenamefont {Pumir},\ and\ \citenamefont {Bodenschatz}}]{BPB2022}%
  \BibitemOpen
  \bibfield  {author} {\bibinfo {author} {\bibfnamefont {D.}~\bibnamefont
  {Buaria}}, \bibinfo {author} {\bibfnamefont {A.}~\bibnamefont {Pumir}},\ and\
  \bibinfo {author} {\bibfnamefont {E.}~\bibnamefont {Bodenschatz}},\
  }\bibfield  {title} {\bibinfo {title} {Generation of intense dissipation in
  high reynolds number turbulence},\ }\href@noop {} {\bibfield  {journal}
  {\bibinfo  {journal} {Phil.~Trans.~R.~Soc.~A}\ }\textbf {\bibinfo {volume}
  {380}},\ \bibinfo {pages} {20210088} (\bibinfo {year} {2022})}\BibitemShut
  {NoStop}%
\bibitem [{\citenamefont {Buaria}\ \emph
  {et~al.}(2020{\natexlab{a}})\citenamefont {Buaria}, \citenamefont {Pumir},\
  and\ \citenamefont {Bodenschatz}}]{BPB2020}%
  \BibitemOpen
  \bibfield  {author} {\bibinfo {author} {\bibfnamefont {D.}~\bibnamefont
  {Buaria}}, \bibinfo {author} {\bibfnamefont {A.}~\bibnamefont {Pumir}},\ and\
  \bibinfo {author} {\bibfnamefont {E.}~\bibnamefont {Bodenschatz}},\
  }\bibfield  {title} {\bibinfo {title} {Self-attenuation of extreme events in
  {Navier-Stokes} turbulence},\ }\href@noop {} {\bibfield  {journal} {\bibinfo
  {journal} {Nat. Commun.}\ }\textbf {\bibinfo {volume} {11}},\ \bibinfo
  {pages} {5852} (\bibinfo {year} {2020}{\natexlab{a}})}\BibitemShut {NoStop}%
\bibitem [{\citenamefont {Ashurst}\ \emph {et~al.}(1987)\citenamefont
  {Ashurst}, \citenamefont {Kerstein}, \citenamefont {Kerr},\ and\
  \citenamefont {Gibson}}]{Ashurst87}%
  \BibitemOpen
  \bibfield  {author} {\bibinfo {author} {\bibfnamefont {W.~T.}\ \bibnamefont
  {Ashurst}}, \bibinfo {author} {\bibfnamefont {A.~R.}\ \bibnamefont
  {Kerstein}}, \bibinfo {author} {\bibfnamefont {R.~M.}\ \bibnamefont {Kerr}},\
  and\ \bibinfo {author} {\bibfnamefont {C.~H.}\ \bibnamefont {Gibson}},\
  }\bibfield  {title} {\bibinfo {title} {Alignment of vorticity and scalar
  gradient with strain rate in simulated {Navier-Stokes} turbulence},\
  }\href@noop {} {\bibfield  {journal} {\bibinfo  {journal} {Phys. Fluids}\
  }\textbf {\bibinfo {volume} {30}},\ \bibinfo {pages} {2343} (\bibinfo {year}
  {1987})}\BibitemShut {NoStop}%
\bibitem [{\citenamefont {Buaria}\ \emph
  {et~al.}(2020{\natexlab{b}})\citenamefont {Buaria}, \citenamefont
  {Bodenschatz},\ and\ \citenamefont {Pumir}}]{BBP2020}%
  \BibitemOpen
  \bibfield  {author} {\bibinfo {author} {\bibfnamefont {D.}~\bibnamefont
  {Buaria}}, \bibinfo {author} {\bibfnamefont {E.}~\bibnamefont
  {Bodenschatz}},\ and\ \bibinfo {author} {\bibfnamefont {A.}~\bibnamefont
  {Pumir}},\ }\bibfield  {title} {\bibinfo {title} {Vortex stretching and
  enstrophy production in high {Reynolds} number turbulence},\ }\href@noop {}
  {\bibfield  {journal} {\bibinfo  {journal} {Phys.~Rev.~Fluids}\ }\textbf
  {\bibinfo {volume} {5}},\ \bibinfo {pages} {104602} (\bibinfo {year}
  {2020}{\natexlab{b}})}\BibitemShut {NoStop}%
\bibitem [{\citenamefont {Tsinober}\ \emph {et~al.}(1999)\citenamefont
  {Tsinober}, \citenamefont {Ortenberg},\ and\ \citenamefont
  {Shtilman}}]{tsi99}%
  \BibitemOpen
  \bibfield  {author} {\bibinfo {author} {\bibfnamefont {A.}~\bibnamefont
  {Tsinober}}, \bibinfo {author} {\bibfnamefont {M.}~\bibnamefont
  {Ortenberg}},\ and\ \bibinfo {author} {\bibfnamefont {L.}~\bibnamefont
  {Shtilman}},\ }\bibfield  {title} {\bibinfo {title} {On depression of
  nonlinearity in turbulence},\ }\href@noop {} {\bibfield  {journal} {\bibinfo
  {journal} {Phys. Fluids}\ }\textbf {\bibinfo {volume} {11}},\ \bibinfo
  {pages} {2291} (\bibinfo {year} {1999})}\BibitemShut {NoStop}%
\bibitem [{\citenamefont {Jim{\'e}nez}\ \emph {et~al.}(1993)\citenamefont
  {Jim{\'e}nez}, \citenamefont {Wray}, \citenamefont {Saffman},\ and\
  \citenamefont {Rogallo}}]{Jimenez93}%
  \BibitemOpen
  \bibfield  {author} {\bibinfo {author} {\bibfnamefont {J.}~\bibnamefont
  {Jim{\'e}nez}}, \bibinfo {author} {\bibfnamefont {A.~A.}\ \bibnamefont
  {Wray}}, \bibinfo {author} {\bibfnamefont {P.~G.}\ \bibnamefont {Saffman}},\
  and\ \bibinfo {author} {\bibfnamefont {R.~S.}\ \bibnamefont {Rogallo}},\
  }\bibfield  {title} {\bibinfo {title} {The structure of intense vorticity in
  isotropic turbulence},\ }\href@noop {} {\bibfield  {journal} {\bibinfo
  {journal} {J. Fluid Mech.}\ }\textbf {\bibinfo {volume} {255}} (\bibinfo
  {year} {1993})}\BibitemShut {NoStop}%
\bibitem [{\citenamefont {Moisy}\ and\ \citenamefont
  {Jim{\'e}nez}(2004)}]{moisy:2004}%
  \BibitemOpen
  \bibfield  {author} {\bibinfo {author} {\bibfnamefont {F.}~\bibnamefont
  {Moisy}}\ and\ \bibinfo {author} {\bibfnamefont {J.}~\bibnamefont
  {Jim{\'e}nez}},\ }\bibfield  {title} {\bibinfo {title} {Geometry and
  clustering of intense structures in isotropic turbulence},\ }\href@noop {}
  {\bibfield  {journal} {\bibinfo  {journal} {J.~Fluid Mech.}\ }\textbf
  {\bibinfo {volume} {513}},\ \bibinfo {pages} {111} (\bibinfo {year}
  {2004})}\BibitemShut {NoStop}%
\bibitem [{\citenamefont {Girimaji}\ and\ \citenamefont
  {Pope}(1989)}]{girimaji:1990}%
  \BibitemOpen
  \bibfield  {author} {\bibinfo {author} {\bibfnamefont {S.~S.}\ \bibnamefont
  {Girimaji}}\ and\ \bibinfo {author} {\bibfnamefont {S.~B.}\ \bibnamefont
  {Pope}},\ }\bibfield  {title} {\bibinfo {title} {A diffusion model for
  velocity gradients in turbulence},\ }\href@noop {} {\bibfield  {journal}
  {\bibinfo  {journal} {Phys.~Fluids A}\ }\textbf {\bibinfo {volume} {2}},\
  \bibinfo {pages} {242} (\bibinfo {year} {1989})}\BibitemShut {NoStop}%
\bibitem [{\citenamefont {Johnson}\ and\ \citenamefont
  {Wilczek}(2023)}]{JM.arfm}%
  \BibitemOpen
  \bibfield  {author} {\bibinfo {author} {\bibfnamefont {P.~L.}\ \bibnamefont
  {Johnson}}\ and\ \bibinfo {author} {\bibfnamefont {M.}~\bibnamefont
  {Wilczek}},\ }\bibfield  {title} {\bibinfo {title} {Multiscale velocity
  gradients in turbulence},\ }\href@noop {} {\bibfield  {journal} {\bibinfo
  {journal} {Annu.~Rev.~Fluid Mech.}\ ,\ \bibinfo {pages} {(to appear)}}
  (\bibinfo {year} {2023})}\BibitemShut {NoStop}%
\bibitem [{\citenamefont {Tian}\ \emph {et~al.}(2021)\citenamefont {Tian},
  \citenamefont {Livescu},\ and\ \citenamefont {Chertkov}}]{tian21}%
  \BibitemOpen
  \bibfield  {author} {\bibinfo {author} {\bibfnamefont {Y.}~\bibnamefont
  {Tian}}, \bibinfo {author} {\bibfnamefont {D.}~\bibnamefont {Livescu}},\ and\
  \bibinfo {author} {\bibfnamefont {M.}~\bibnamefont {Chertkov}},\ }\bibfield
  {title} {\bibinfo {title} {Physics-informed machine learning of the
  {Lagrangian} dynamics of velocity gradient tensor},\ }\href@noop {}
  {\bibfield  {journal} {\bibinfo  {journal} {Phys.~Rev.~Fluids}\ }\textbf
  {\bibinfo {volume} {6}},\ \bibinfo {pages} {094607} (\bibinfo {year}
  {2021})}\BibitemShut {NoStop}%
\bibitem [{\citenamefont {Buaria}\ and\ \citenamefont
  {Sreenivasan}(2023{\natexlab{a}})}]{bs_pnas_23}%
  \BibitemOpen
  \bibfield  {author} {\bibinfo {author} {\bibfnamefont {D.}~\bibnamefont
  {Buaria}}\ and\ \bibinfo {author} {\bibfnamefont {K.~R.}\ \bibnamefont
  {Sreenivasan}},\ }\bibfield  {title} {\bibinfo {title} {Forecasting small
  scale dynamics of fluid turbulence using deep neural networks},\ }\href@noop
  {} {\bibfield  {journal} {\bibinfo  {journal} {Proc. Nat. Acad. Sci.}\
  }\textbf {\bibinfo {volume} {{120}}},\ \bibinfo {pages} {{e2305765120}}
  (\bibinfo {year} {2023}{\natexlab{a}})}\BibitemShut {NoStop}%
\bibitem [{\citenamefont {Hamlington}\ \emph {et~al.}(2008)\citenamefont
  {Hamlington}, \citenamefont {Schumacher},\ and\ \citenamefont
  {Dahm}}]{ham_pof08}%
  \BibitemOpen
  \bibfield  {author} {\bibinfo {author} {\bibfnamefont {P.~E.}\ \bibnamefont
  {Hamlington}}, \bibinfo {author} {\bibfnamefont {J.}~\bibnamefont
  {Schumacher}},\ and\ \bibinfo {author} {\bibfnamefont {W.~J.~A.}\
  \bibnamefont {Dahm}},\ }\bibfield  {title} {\bibinfo {title} {Direct
  assessment of vorticity alignment with local and nonlocal strain rates in
  turbulent flows},\ }\href@noop {} {\bibfield  {journal} {\bibinfo  {journal}
  {Phys.~Fluids}\ }\textbf {\bibinfo {volume} {20}},\ \bibinfo {pages} {111703}
  (\bibinfo {year} {2008})}\BibitemShut {NoStop}%
\bibitem [{\citenamefont {Buaria}\ and\ \citenamefont {Pumir}(2021)}]{BP2021}%
  \BibitemOpen
  \bibfield  {author} {\bibinfo {author} {\bibfnamefont {D.}~\bibnamefont
  {Buaria}}\ and\ \bibinfo {author} {\bibfnamefont {A.}~\bibnamefont {Pumir}},\
  }\bibfield  {title} {\bibinfo {title} {Nonlocal amplification of intense
  vorticity in turbulent flows},\ }\href@noop {} {\bibfield  {journal}
  {\bibinfo  {journal} {Phys. Rev. Research}\ }\textbf {\bibinfo {volume}
  {3}},\ \bibinfo {pages} {L042020} (\bibinfo {year} {2021})}\BibitemShut
  {NoStop}%
\bibitem [{\citenamefont {Vlaykov}\ and\ \citenamefont
  {Wilczek}(2019)}]{Vlaykov2019Small}%
  \BibitemOpen
  \bibfield  {author} {\bibinfo {author} {\bibfnamefont {D.~G.}\ \bibnamefont
  {Vlaykov}}\ and\ \bibinfo {author} {\bibfnamefont {M.}~\bibnamefont
  {Wilczek}},\ }\bibfield  {title} {\bibinfo {title} {On the small-scale
  structure of turbulence and its impact on the pressure field},\ }\href@noop
  {} {\bibfield  {journal} {\bibinfo  {journal} {J.~Fluid Mech.}\ }\textbf
  {\bibinfo {volume} {861}},\ \bibinfo {pages} {422–446} (\bibinfo {year}
  {2019})}\BibitemShut {NoStop}%
\bibitem [{\citenamefont {Buaria}\ \emph {et~al.}(2019)\citenamefont {Buaria},
  \citenamefont {Pumir}, \citenamefont {Bodenschatz},\ and\ \citenamefont
  {Yeung}}]{BPBY2019}%
  \BibitemOpen
  \bibfield  {author} {\bibinfo {author} {\bibfnamefont {D.}~\bibnamefont
  {Buaria}}, \bibinfo {author} {\bibfnamefont {A.}~\bibnamefont {Pumir}},
  \bibinfo {author} {\bibfnamefont {E.}~\bibnamefont {Bodenschatz}},\ and\
  \bibinfo {author} {\bibfnamefont {P.~K.}\ \bibnamefont {Yeung}},\ }\bibfield
  {title} {\bibinfo {title} {Extreme velocity gradients in turbulent flows},\
  }\href@noop {} {\bibfield  {journal} {\bibinfo  {journal} {New J.~Phys.}\
  }\textbf {\bibinfo {volume} {21}},\ \bibinfo {pages} {043004} (\bibinfo
  {year} {2019})}\BibitemShut {NoStop}%
\bibitem [{\citenamefont {Buaria}\ and\ \citenamefont {Pumir}(2022)}]{BP2022}%
  \BibitemOpen
  \bibfield  {author} {\bibinfo {author} {\bibfnamefont {D.}~\bibnamefont
  {Buaria}}\ and\ \bibinfo {author} {\bibfnamefont {A.}~\bibnamefont {Pumir}},\
  }\bibfield  {title} {\bibinfo {title} {Vorticity-strain rate dynamics and the
  smallest scales of turbulence},\ }\href@noop {} {\bibfield  {journal}
  {\bibinfo  {journal} {Phys. Rev. Lett.}\ }\textbf {\bibinfo {volume} {128}},\
  \bibinfo {pages} {094501} (\bibinfo {year} {2022})}\BibitemShut {NoStop}%
\bibitem [{\citenamefont {Ohkitani}\ and\ \citenamefont
  {Kishiba}(1995)}]{Ohkitani:95}%
  \BibitemOpen
  \bibfield  {author} {\bibinfo {author} {\bibfnamefont {K.}~\bibnamefont
  {Ohkitani}}\ and\ \bibinfo {author} {\bibfnamefont {S.}~\bibnamefont
  {Kishiba}},\ }\bibfield  {title} {\bibinfo {title} {Nonlocal nature of vortex
  stretching in an inviscid fluid},\ }\href@noop {} {\bibfield  {journal}
  {\bibinfo  {journal} {Phys.~Fluids}\ }\textbf {\bibinfo {volume} {7}},\
  \bibinfo {pages} {411} (\bibinfo {year} {1995})}\BibitemShut {NoStop}%
\bibitem [{\citenamefont {Buaria}\ and\ \citenamefont
  {Sreenivasan}(2020)}]{BS2020}%
  \BibitemOpen
  \bibfield  {author} {\bibinfo {author} {\bibfnamefont {D.}~\bibnamefont
  {Buaria}}\ and\ \bibinfo {author} {\bibfnamefont {K.~R.}\ \bibnamefont
  {Sreenivasan}},\ }\bibfield  {title} {\bibinfo {title} {Dissipation range of
  the energy spectrum in high {Reynolds} number turbulence},\ }\href@noop {}
  {\bibfield  {journal} {\bibinfo  {journal} {Phys.~Rev.~Fluids}\ }\textbf
  {\bibinfo {volume} {5}},\ \bibinfo {pages} {092601(R)} (\bibinfo {year}
  {2020})}\BibitemShut {NoStop}%
\bibitem [{\citenamefont {Buaria}\ and\ \citenamefont
  {Sreenivasan}(2022{\natexlab{a}})}]{BS2022}%
  \BibitemOpen
  \bibfield  {author} {\bibinfo {author} {\bibfnamefont {D.}~\bibnamefont
  {Buaria}}\ and\ \bibinfo {author} {\bibfnamefont {K.~R.}\ \bibnamefont
  {Sreenivasan}},\ }\bibfield  {title} {\bibinfo {title} {Intermittency of
  turbulent velocity and scalar fields using three-dimensional local
  averaging},\ }\href@noop {} {\bibfield  {journal} {\bibinfo  {journal}
  {Phys.~Rev.~Fluids}\ }\textbf {\bibinfo {volume} {7}},\ \bibinfo {pages}
  {L072601} (\bibinfo {year} {2022}{\natexlab{a}})}\BibitemShut {NoStop}%
\bibitem [{\citenamefont {Buaria}\ and\ \citenamefont
  {Sreenivasan}(2023{\natexlab{b}})}]{BS2023}%
  \BibitemOpen
  \bibfield  {author} {\bibinfo {author} {\bibfnamefont {D.}~\bibnamefont
  {Buaria}}\ and\ \bibinfo {author} {\bibfnamefont {K.~R.}\ \bibnamefont
  {Sreenivasan}},\ }\bibfield  {title} {\bibinfo {title} {Lagrangian
  acceleration and its {Eulerian} decompositions in fully developed
  turbulence},\ }\href@noop {} {\bibfield  {journal} {\bibinfo  {journal}
  {Phys.~Rev.~Fluids}\ }\textbf {\bibinfo {volume} {8}},\ \bibinfo {pages}
  {L032601} (\bibinfo {year} {2023}{\natexlab{b}})}\BibitemShut {NoStop}%
\bibitem [{\citenamefont {Rogallo}(1981)}]{Rogallo}%
  \BibitemOpen
  \bibfield  {author} {\bibinfo {author} {\bibfnamefont {R.~S.}\ \bibnamefont
  {Rogallo}},\ }\bibfield  {title} {\bibinfo {title} {Numerical experiments in
  homogeneous turbulence},\ }\href@noop {} {\bibfield  {journal} {\bibinfo
  {journal} {NASA Technical Memo}\ }\textbf {\bibinfo {volume} {81315}}
  (\bibinfo {year} {1981})}\BibitemShut {NoStop}%
\bibitem [{\citenamefont {Batchelor}(1953)}]{batchelor53}%
  \BibitemOpen
  \bibfield  {author} {\bibinfo {author} {\bibfnamefont {G.~K.}\ \bibnamefont
  {Batchelor}},\ }\href@noop {} {\emph {\bibinfo {title} {The theory of
  homogeneous turbulence}}}\ (\bibinfo  {publisher} {Cambridge university
  press},\ \bibinfo {year} {1953})\BibitemShut {NoStop}%
\bibitem [{\citenamefont {Betchov}(1956)}]{Betchov56}%
  \BibitemOpen
  \bibfield  {author} {\bibinfo {author} {\bibfnamefont {R.}~\bibnamefont
  {Betchov}},\ }\bibfield  {title} {\bibinfo {title} {An inequality concerning
  the production of vorticity in isotropic turbulence},\ }\href@noop {}
  {\bibfield  {journal} {\bibinfo  {journal} {J.~Fluid Mech.}\ }\textbf
  {\bibinfo {volume} {1}},\ \bibinfo {pages} {497} (\bibinfo {year}
  {1956})}\BibitemShut {NoStop}%
\bibitem [{\citenamefont {Kerr}(1985)}]{kerr85}%
  \BibitemOpen
  \bibfield  {author} {\bibinfo {author} {\bibfnamefont {R.~M.}\ \bibnamefont
  {Kerr}},\ }\bibfield  {title} {\bibinfo {title} {Higher-order derivative
  correlations and the alignment of small-scale structures in isotropic
  numerical turbulence},\ }\href@noop {} {\bibfield  {journal} {\bibinfo
  {journal} {J.~Fluid Mech.}\ }\textbf {\bibinfo {volume} {153}},\ \bibinfo
  {pages} {31} (\bibinfo {year} {1985})}\BibitemShut {NoStop}%
\bibitem [{\citenamefont {Gylfason}\ \emph {et~al.}(2004)\citenamefont
  {Gylfason}, \citenamefont {Ayyalasomayjula},\ and\ \citenamefont
  {Warhaft}}]{gylfason:2004}%
  \BibitemOpen
  \bibfield  {author} {\bibinfo {author} {\bibfnamefont {A.}~\bibnamefont
  {Gylfason}}, \bibinfo {author} {\bibfnamefont {S.}~\bibnamefont
  {Ayyalasomayjula}},\ and\ \bibinfo {author} {\bibfnamefont {Z.}~\bibnamefont
  {Warhaft}},\ }\bibfield  {title} {\bibinfo {title} {Intermittency, pressure
  and acceleration statistics from hot-wire measurements in wind-tunnel
  turbulence},\ }\href@noop {} {\bibfield  {journal} {\bibinfo  {journal}
  {J.~Fluid Mech.}\ }\textbf {\bibinfo {volume} {501}},\ \bibinfo {pages} {213}
  (\bibinfo {year} {2004})}\BibitemShut {NoStop}%
\bibitem [{\citenamefont {Buaria}\ and\ \citenamefont
  {Sreenivasan}(2022{\natexlab{b}})}]{BS_PRL_2022}%
  \BibitemOpen
  \bibfield  {author} {\bibinfo {author} {\bibfnamefont {D.}~\bibnamefont
  {Buaria}}\ and\ \bibinfo {author} {\bibfnamefont {K.~R.}\ \bibnamefont
  {Sreenivasan}},\ }\bibfield  {title} {\bibinfo {title} {Scaling of
  acceleration statistics in high {Reynolds} number turbulence},\ }\href@noop
  {} {\bibfield  {journal} {\bibinfo  {journal} {Phys.~Rev.~Lett.}\ }\textbf
  {\bibinfo {volume} {128}},\ \bibinfo {pages} {234502} (\bibinfo {year}
  {2022}{\natexlab{b}})}\BibitemShut {NoStop}%
\bibitem [{\citenamefont {Ishihara}\ \emph {et~al.}(2007)\citenamefont
  {Ishihara}, \citenamefont {Kaneda}, \citenamefont {Yokokawa}, \citenamefont
  {Itakura},\ and\ \citenamefont {Uno}}]{Ishihara07}%
  \BibitemOpen
  \bibfield  {author} {\bibinfo {author} {\bibfnamefont {T.}~\bibnamefont
  {Ishihara}}, \bibinfo {author} {\bibfnamefont {Y.}~\bibnamefont {Kaneda}},
  \bibinfo {author} {\bibfnamefont {M.}~\bibnamefont {Yokokawa}}, \bibinfo
  {author} {\bibfnamefont {K.}~\bibnamefont {Itakura}},\ and\ \bibinfo {author}
  {\bibfnamefont {A.}~\bibnamefont {Uno}},\ }\bibfield  {title} {\bibinfo
  {title} {Small-scale statistics in high resolution of numerically isotropic
  turbulence},\ }\href@noop {} {\bibfield  {journal} {\bibinfo  {journal}
  {J.~Fluid~Mech.}\ }\textbf {\bibinfo {volume} {592}},\ \bibinfo {pages} {335}
  (\bibinfo {year} {2007})}\BibitemShut {NoStop}%
\bibitem [{\citenamefont {Burgers}(1948)}]{Burgers48}%
  \BibitemOpen
  \bibfield  {author} {\bibinfo {author} {\bibfnamefont {J.~M.}\ \bibnamefont
  {Burgers}},\ }\bibfield  {title} {\bibinfo {title} {A mathematical model
  illustrating the theory of turbulence},\ }\href@noop {} {\bibfield  {journal}
  {\bibinfo  {journal} {Adv.~Appl.~Mech.}\ }\textbf {\bibinfo {volume} {1}},\
  \bibinfo {pages} {171} (\bibinfo {year} {1948})}\BibitemShut {NoStop}%
\bibitem [{\citenamefont {Andreotti}(1997)}]{andreotti1997}%
  \BibitemOpen
  \bibfield  {author} {\bibinfo {author} {\bibfnamefont {B.}~\bibnamefont
  {Andreotti}},\ }\bibfield  {title} {\bibinfo {title} {{Studying Burgers’
  models to investigate the physical meaning of the alignments statistically
  observed in turbulence}},\ }\href@noop {} {\bibfield  {journal} {\bibinfo
  {journal} {Phys.~Fluids}\ }\textbf {\bibinfo {volume} {9}},\ \bibinfo {pages}
  {735} (\bibinfo {year} {1997})}\BibitemShut {NoStop}%
\bibitem [{\citenamefont {Choi}\ \emph {et~al.}(2009)\citenamefont {Choi},
  \citenamefont {Kim},\ and\ \citenamefont {Lee}}]{Choi:09}%
  \BibitemOpen
  \bibfield  {author} {\bibinfo {author} {\bibfnamefont {Y.}~\bibnamefont
  {Choi}}, \bibinfo {author} {\bibfnamefont {B.~G.}\ \bibnamefont {Kim}},\ and\
  \bibinfo {author} {\bibfnamefont {C.}~\bibnamefont {Lee}},\ }\bibfield
  {title} {\bibinfo {title} {Alignment of velocity and vorticity and the
  intermittent distribution of helicity in isotropic turbulence},\ }\href@noop
  {} {\bibfield  {journal} {\bibinfo  {journal} {Phys. Rev. E}\ }\textbf
  {\bibinfo {volume} {80}},\ \bibinfo {pages} {017301} (\bibinfo {year}
  {2009})}\BibitemShut {NoStop}%
\bibitem [{\citenamefont {Pope}(2000)}]{popebook}%
  \BibitemOpen
  \bibfield  {author} {\bibinfo {author} {\bibfnamefont {S.~B.}\ \bibnamefont
  {Pope}},\ }\href@noop {} {\emph {\bibinfo {title} {Turbulent Flows}}}\
  (\bibinfo  {publisher} {Cambridge University Press},\ \bibinfo {year}
  {2000})\BibitemShut {NoStop}%
\end{thebibliography}
\end{document}